\documentclass[aps,pra,superscriptaddress,reprint,floatfix,longbibliography,showpacs]{revtex4-1}
\usepackage[english]{babel}
\usepackage{amsmath}
\usepackage{amssymb}
\usepackage{graphicx}
\usepackage{hyperref}
\usepackage{upgreek}
\usepackage{epstopdf}
\usepackage{color}
\usepackage{dcolumn}
\usepackage{multirow}
\usepackage[note-name=, use-sort-key = false]{notes2bib}

\begin{document}
\title{Optical signature of the pressure-induced dimerization in the \\
honeycomb iridate $\upalpha$-Li$_2$IrO$_3$}

\author{V. Hermann}
\affiliation{Experimentalphysik II, University of Augsburg, 86159 Augsburg, Germany}
\author{J. Ebad-Allah}
\affiliation{Experimentalphysik II, University of Augsburg, 86159 Augsburg, Germany}
\affiliation{Department of Physics, Tanta University, 31527 Tanta, Egypt}
\author{F.~Freund}
\author{A.~Jesche}
\author{A.~A.~Tsirlin}
\affiliation{Experimentalphysik VI, Center for Electronic Correlations and Magnetism, University of Augsburg, 86159 Augsburg, Germany}
\author{P. Gegenwart}
\affiliation{Experimentalphysik VI, Center for Electronic Correlations and Magnetism, University of Augsburg, 86159 Augsburg, Germany}
\author{C. A. Kuntscher}
\affiliation{Experimentalphysik II, University of Augsburg, 86159 Augsburg, Germany}
\email{christine.kuntscher@physik.uni-augsburg.de}

\begin{abstract}
We studied the effect of external pressure on the electrodynamic properties of $\upalpha$-Li\textsubscript{2}IrO\textsubscript{3} single crystals in the frequency range of the phonon modes and the Ir $d$-$d$ transitions.
The abrupt hardening of several phonon modes under pressure supports the onset of the dimerized phase at the critical pressure $P\textsubscript{c}$=3.8~GPa.
With increasing pressure an overall decrease in spectral weight of the Ir $d$-$d$ transitions is found up to $P\textsubscript{c}$.
Above $P\textsubscript{c}$, the local (on-site) $d$-$d$ excitations gain spectral weight with increasing pressure, which hints at a pressure-induced increase in the octahedral distortions.
The non-local (intersite) Ir $d$-$d$ transitions show a monotonic blue-shift and decrease in spectral weight. The changes observed for the non-local excitations are most prominent well above $P\textsubscript{c}$, namely for pressures $\geq$12~GPa, and only small changes occur for pressures close to $P\textsubscript{c}$. The profile of the optical conductivity at high pressures ($\sim$20~GPa) appears to be indicative for the dimerized state in iridates.
\end{abstract}

\maketitle

\section{Introduction}

In recent years layered honeycomb systems such as Na$_2$IrO$_3$, $\upalpha$-Li$_2$IrO$_3$, and $\upalpha$-RuCl$_3$ attracted the condensed matter community due to the possible realization of the Kitaev quantum spin-liquid (QSL) ground state~\cite{Kitaev.2006, Jackeli.2009, Chaloupka.2010,Singh.2012, Choi.2012, Plumb.2014, Chun.2015, Winter.2017}.
However, these compounds were found to order magnetically at low temperatures~\cite{Choi.2012, Sears.2015, Banerjee.2016, Williams.2016} excluding the realization of this desired quantum state at ambient conditions, as not only anisotropic Kitaev but also isotropic Heisenberg and other types of interactions are present in these materials.
Effort was taken to tune the most promising candidate $\upalpha$-RuCl$_3$ towards the QSL by high external magnetic field~\cite{Wolter.2017,Baek.2017}, but the results are strongly debated until now~\cite{Winter.2017,Janssen.2016,Winter.2018,Takagi.2019}.

A different approach for the realization of the Kitaev QSL is the application of external or chemical pressure on the honeycomb lattice~\cite{Hermann.2017,Biesner.2018,Simutis.2018}, since the hopping parameters could possibly be changed in favor of the Kitaev over Heisenberg interactions.
However, it was shown that the Neel temperature of the compounds Na$_2$IrO$_3$ and $\upalpha$-RuCl$_3$ {\it increases} under low external pressure
probably due to the increase in interlayer interaction~\cite{Simutis.2018,Wang.2018}, as hydrostatic pressure mainly compresses the interlayer distance in these materials~\cite{Hermann.2017,Hermann.2018}.
Furthermore, the QSL ground state is in competition with a dimerized ground state.
It was found that the iridates as well as $\upalpha$-RuCl$_3$ all dimerize at finite pressures~\cite{Hermann.2018,Biesner.2018,Bastien.2018,Hu.2018,Majumder.2018}, whereby the critical pressure for dimerization depends on several parameters, such as the type of stabilizing central ion, the strength of spin-orbit coupling, and electronic correlations~\cite{Hermann.2018}
\bibnote{For Na$_2$IrO$_3$, additional pressure-induced transitions were suggested recently from powder x-ray diffraction measurements~\cite{Xi.2018}, which were, however, not observed by higher-quality single-crystal x-ray diffraction data~\cite{Hermann.2017}.}.
Apparently, hydrostatic pressure is not the most promising approach to tune the honeycomb materials towards the desired Kitaev QSL, and it was concluded that, instead, strain would induce the QSL state~\cite{Yadav.2018}, for example by chemical pressure~\cite{Hermann.2017,Simutis.2018}.
Nevertheless, hydrostatic external pressure is one of the superior ways to tune the electronic ground state of these systems.

The layered honeycomb 213 iridates Na\textsubscript{2}IrO\textsubscript{3} and $\upalpha$-Li\textsubscript{2}IrO\textsubscript{3} contain Ir$^{4+}$ ions with a $d^5$ electronic configuration and edge-sharing IrO\textsubscript{6} octahedra. The electronic ground state is intensively debated due to the interplay of Coulomb repulsion $U$, Hund's rule coupling $J_H$, spin-orbit coupling $\lambda$, crystal field splitting, and electronic bandwidth (given by the hopping parameter $t$), which are all of similar energy scale \cite{Gretarsson.2013}.
Depending on the hierarchy of these parameters, the 213 iridates were theoretically discussed in terms of semimetals~\cite{Kim.2012,Kim.2013}, relativistic Mott insulators~\cite{Jackeli.2009}, or within an itinerant picture with the formation of quasi-molecular orbitals (QMOs)~\cite{Mazin.2012,Mazin.2013}.
Experiments finally showed that the electronic ground state of Na\textsubscript{2}IrO\textsubscript{3} can be well described in terms of a relativistic Mott insulator with a half-filled $j\textsubscript{eff}$=$1/2$-band split by the Hubbard $U$~\cite{Singh.2010,Gretarsson.2013,Hermann.2017,Xi.2018}, although it exhibits a significant intrahexagon itinerant character which could lead to a mixing of
$j\textsubscript{eff}$=$1/2$ ($j_{1/2}$) and $j\textsubscript{eff}$=$3/2$ ($j_{3/2}$) states in the Ir t\textsubscript{2g} level~\cite{Sohn.2013,Kim.2016c,Hermann.2017}.
Compared to Na\textsubscript{2}IrO\textsubscript{3}, for $\upalpha$-Li\textsubscript{2}IrO\textsubscript{3} signs for an increased QMO character was observed at ambient conditions~\cite{Hermann.2017} and at low pressures~\cite{Clancy.2018}.

Two hopping paths are most important for the realization of the QMO ground state in the iridates~\cite{Mazin.2012}: i) The direct metal-metal hopping between two neighboring Ir ions, and ii) the indirect and spatially anisotropic hopping between two neighboring Ir $t\textsubscript{2g}$ orbitals mediated by the O 2p orbitals.
The former depends on the Ir--Ir bond length, while the latter depends on the Ir--O bond lengths as well as the Ir--O--Ir bond angles.
Small distortions of the Ir--Ir network and the IrO\textsubscript{6} octahedra are proposed to favor the formation of QMOs~\cite{Foyevtsova.2013}.
Strong distortions of the Ir--Ir network, however, will probably prevent the realization of a pure QMO picture, and strong trigonal distortions of the octahedra will lead to a breakdown of the relativistic Mott states~\cite{Bhattacharjee.2012}.
However, the effect of external pressure on electronic states can be more complicated, as was recently shown by studies on Na$_3$Ir$_3$O$_8$, which exhibits a 1/3-filled $j_{1/2}$-band and is therefore semimetallic at ambient pressure~\cite{Sun.2018}.
By applying external pressure, the $j_{1/2}$-band in Na$_3$Ir$_3$O$_8$ gets closer to half-filling due to charge transfer introduced by the stronger overlap of the orbitals, which induces an unusual semimetal to insulator transition~\cite{Sun.2018}.

In our previous single-crystal x-ray diffraction (XRD) study on $\upalpha$-Li$_2$IrO$_3$ we observed a pressure-induced dimerization of two of the six Ir--Ir bonds per hexagon at $P\textsubscript{c}$=3.8\,GPa due to a subtle interplay between
magnetism, electronic correlation, spin-orbit coupling, and covalent bonding~\cite{Hermann.2018}. In contrast to the recent study of Clancy \textit{et al.}\cite{Clancy.2018}, where structural and electronic changes were found right above ambient pressure, i.e., well below $P\textsubscript{c}$, we did not find clear indications for structural changes below the sharp transition at $P\textsubscript{c}$ \cite{Hermann.2018}.
Here, we study the effect of external pressure on the electrodynamic properties of $\upalpha$-Li\textsubscript{2}IrO\textsubscript{3} single crystals. One focus of this study lies on the changes in the phonon mode spectrum at the pressure-induced dimerization of the Ir--Ir bonds at $P\textsubscript{c}$.
Possible pressure-induced structural changes below $P\textsubscript{c}$ should affect the phonon mode spectrum, from which the discrepancy between our previous XRD study and the study of Clancy \textit{et al.}~\cite{Clancy.2018} could be resolved.
The second focus lies on the energy range of the on-site and intersite Ir $d$-$d$ excitations, which depend on the interplay of the various hopping parameters and the crystal field effects.
The pressure-dependent behavior of the Ir $d$-$d$ transitions give insight into the electronic ground state of $\upalpha$-Li$_2$IrO$_3$ under pressure, which is not yet clear especially in the dimerized phase~\cite{Clancy.2018}.

\section{Methods}

$\upalpha$-Li$_2$IrO$_3$ single crystals were grown by vapor transport of separated educts as described in Ref. \cite{Freund.2016} using elemental lithium and iridium as starting materials.
The samples were characterized by x-ray diffraction, specific heat, and magnetic susceptibility measurements in order to ensure phase-purity and crystal quality. No foreign phases were detected, and a sharp magnetic transition at 15\,K was observed.

Pressure-dependent reflectance and transmission measurements were carried out with a Bruker Vertex v80 Fourier transform infrared spectrometer in combination with an infrared microscope (Bruker Hyperion) equipped with a 15$\times$ Cassegrain objective. All measurements were done at room temperature.
For the measurements in the frequency ranges 150-700\,cm$^{-1}$  and 2000-10000\,cm$^{-1}$ quasi-hydrostatic pressures up to 8.4 and 14.3~GPa, respectively, were achieved by using commercial Diacell CryoDAC-Mega (almax-easylab) diamond-anvil cells (DACs).
For the measurements in the frequency range 2000-18500\,cm$^{-1}$ a custom-made Syassen-Holzapfel-type \cite{Huber.1977} DAC was used to achieve pressures up to 24.4\,GPa. CsI powder served as quasihydrostatic pressure-transmitting medium. The pressure was determined \textit{in situ} using the ruby-luminescence technique \cite{Mao.1986}.

The pressure-dependent reflectivity spectrum $R_\text{s-d}(\omega)$ was calculated according to $R_\text{s-d}(\omega)$=$R_\text{dia} I_\text{s}(\omega)/I_\text{ref,dia}(\omega)$,
where $I_\text{s}$ is the intensity of the radiation reflected from the sample-diamond interface, $I_\text{ref,dia}$ is the intensity reflected from the inner diamond-air interface of the empty DAC, and $R_\text{dia}$ is the reflectivity of diamond, $R_\text{dia}$=0.167, which was assumed to be independent of pressure~\cite{Eremets.1992,Ruoff.1994}.
The reflectivity spectra $R_\text{s-d}$ in the near-infrared and visible frequency ranges were calibrated against a simulated $R_\text{s-d}$ spectrum based on a Lorentz fitting of the reflectivity spectrum measured at the sample-air interface.
The pressure-dependent transmission spectrum $T(\omega)$ was calculated according to $T(\omega)=I\textsubscript{s}(\omega)/I\textsubscript{CsI}(\omega)$, where $I\textsubscript{s}$ is the intensity of the radiation transmitted through the thin sample embedded in the pressure-transmitting medium and $I\textsubscript{CsI}$ is the intensity of the radiation transmitted through the pressure-transmitting medium at a sample-free space inside the same pressure cell.

The low-frequency optical conductivity covering the phonon modes was obtained from the reflectivity spectra $R_\text{s-d}(\omega)$ using the variational dielectric function \cite{Kuzmenko.2005}.
In the near-infrared and visible frequency ranges the optical conductivity was obtained by Kramers-Kronig analysis of the reflectivity $R_\text{s-d}$, taking into account the diamond-sample interface \cite{Plaskett.1963}.
The absorbance was calculated from the transmission spectrum according to $A=-\log T$.

\section{Results and Discussion}

\begin{table}[b]
	\caption{Phonon mode frequencies for $\upalpha$-Li$_2$IrO$_3$ at the lowest pressure (0.6~GPa) together with the atoms mainly involved according to Ref.~\cite{Hermann.2017}.}\label{tab.Phonon_Frequencies}
	\begin{ruledtabular}
		\begin{tabular}{ccc}
			Mode & Frequency (cm$^{-1}$) & Atoms \\\hline
			1 & 573 & O \\
			2 & 539 & Ir--O--Li \\
			3 & 512 & Ir--O--Li \\
			4 & 429 & Li \\
			5 & 407 & Li \\
			6 & 387 & Li \\
			7 & 341 & Li \\
			8 & 220 & Li \\
			9 & 195 & Li \\
		\end{tabular}
	\end{ruledtabular}
\end{table}

\begin{figure}[t]
	\includegraphics[width=0.95\linewidth]{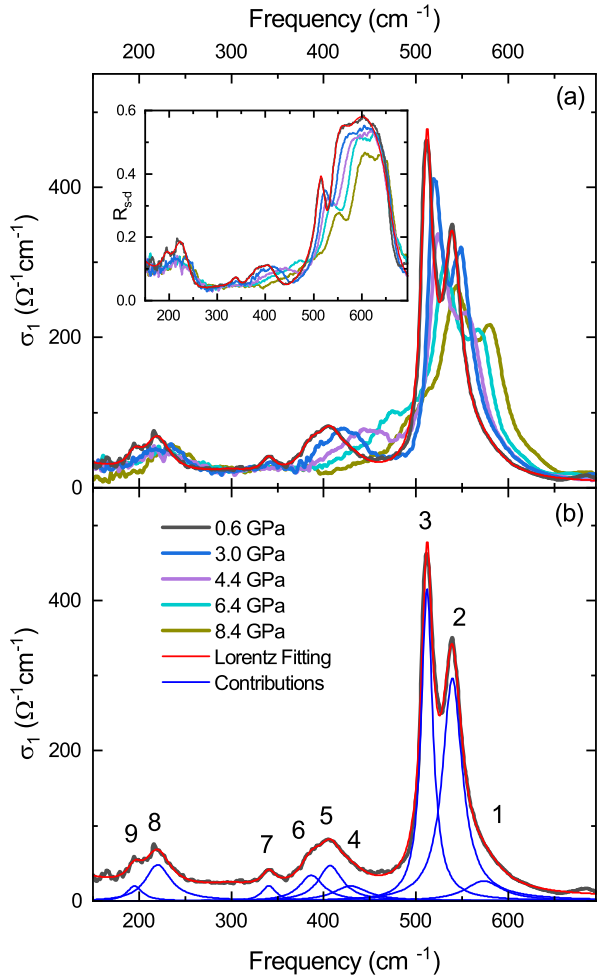}
	\caption{(a) Optical conductivity spectra in the range of the phonon modes for selected pressures. The inset shows the corresponding pressure-dependent reflectivity spectra $R_\text{s-d}$. (b) Optical conductivity spectrum at 0.6~GPa together with the Lorentz fitting (solid red line) and the contributions of the nine experimentally observed phonon modes with their assignment.}\label{fig.Phonon_Modes}
\end{figure}

\subsection{Phonon modes}

The low-frequency optical conductivity spectra of $\upalpha$-Li$_2$IrO$_3$ are depicted in Fig.~\ref{fig.Phonon_Modes}(a) for selected pressures, with the corresponding reflectivity spectra $R_\text{s-d}$ shown in the inset. Both optical conductivity and reflectivity spectra can be well fitted with the Lorentz model, as demonstrated by the Lorentz fittings at the lowest measured pressure (0.6 GPa).
Experimentally nine phonon modes are clearly observed in the frequency range 150-700~cm$^{-1}$, whose contributions to the optical conductivity are shown in Fig.~\ref{fig.Phonon_Modes}(b) for 0.6~GPa.
The frequencies of the modes at 0.6 GPa are listed in Table~\ref{tab.Phonon_Frequencies}, together with the atoms showing the highest displacement (by comparing with the theoretical frequency obtained from Ref.~\cite{Hermann.2017}).
Theoretically, up to 18 infrared-active modes are expected for $\upalpha$-Li$_2$IrO$_3$ due to the $C2/m$ symmetry.
The calculated frequencies of the twelve expected modes in the range 250-700~cm$^{-1}$ are given in Ref.~\cite{Hermann.2017}.
Only seven out of these predicted modes are experimentally well observed, due to the near degeneracy of modes around the modes 2 and 3 and weak modes with low spectral weight between 250-350~cm$^{-1}$.
Two additional modes occur below 250~cm$^{-1}$ (modes 8 and 9), which are also dominated by Li atoms, but were not discussed in detail in Ref.~\cite{Hermann.2017}.

\begin{figure}[t]
	\includegraphics[width=\linewidth]{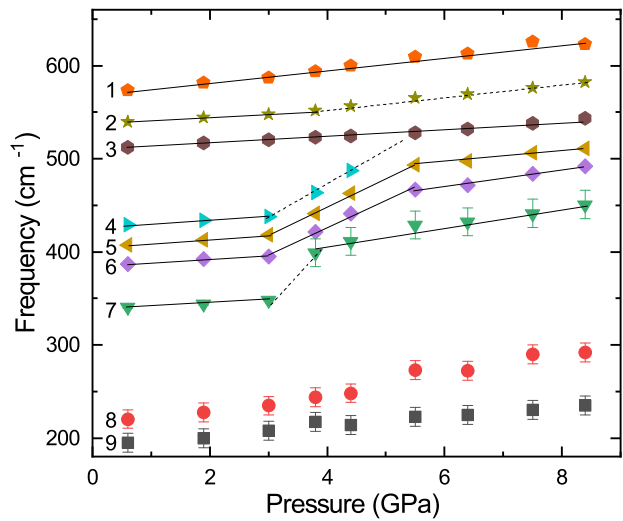}
	\caption{Pressure dependence of the frequencies of the observed nine phonon modes. The solid and dashed lines are guides to the eye.}\label{fig.Phonon_Pressure}
\end{figure}

The behavior of the phonon mode frequencies with increasing pressure is shown in Fig.~\ref{fig.Phonon_Pressure}.
Up to {$P_\text{c}$}$\approx$3.8~GPa all modes harden monotonically, as expected from the pressure-induced decrease in the cell parameters \cite{Hermann.2018}. The smooth pressure evolution of the modes in this pressure range supports the absence of any structural phase transition, in agreement with our single-crystal XRD data \cite{Hermann.2018} and in disagreement with the claim on structural changes at low pressures in Ref.\ \cite{Clancy.2018}.

At $P_\text{c}$ the phonon modes 4-7 show a drastic increase in the slope of their pressure-induced hardening, which in the case of modes 5 and 6 extends up to a pressure of around 5.5~GPa.
Mode 4 shifts into the frequency range of the two strong modes 2 and 3, where it is no longer resolvable, and mode 7 abruptly hardens at $P_\text{c}$.
A significant change of the Ir--O--Li based modes at $P_\text{c}$ is observed only for mode 2, where above the critical pressure the hardening is increased.

From XRD measurements it is known that the Ir--Ir bonds dimerize at a critical pressure of {$P_\text{c}$}$\approx$3.8~GPa~\cite{Hermann.2018}, which furthermore introduces a change of the symmetry from $C2/m$ to $P\overline{1}$.
Also in the dimerized phase, 18 infrared-active modes are expected according to group theory~\cite{Kroumova.2003}.
Interestingly, all the modes, which undergo drastic changes at the onset of dimerization, belong to the group exhibiting mainly Li displacements at ambient pressure.
Accordingly, assuming that the assignment of the phonon modes still holds in the high-pressure, dimerized phase above $P_\text{c}$, the Ir--O bond distances are not strongly affected by the pressure-induced dimerization of the Ir-Ir network.
Significant changes of the edge-sharing Ir--O--Ir bond angles could, however, be induced by pressure, which would then affect the Ir $d$-$d$ transitions located in the higher-energy ranges, as will be discussed in the following.

\subsection{Ir $d$-$d$ transitions}

\begin{figure}[t]
	\includegraphics[width=\linewidth]{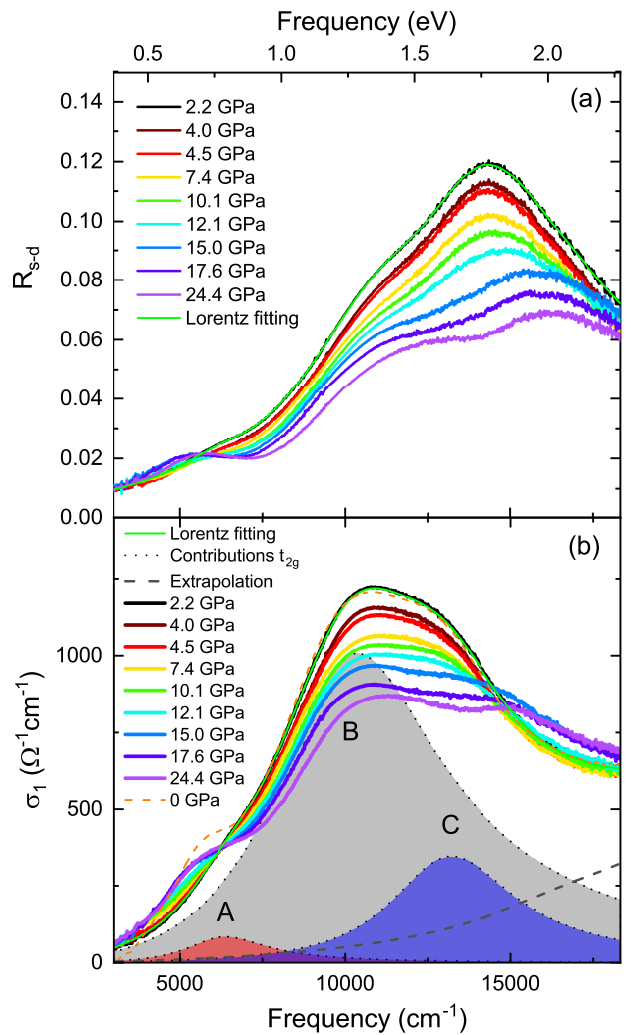}
	\caption{(a) Reflectivity spectra $R_\text{s-d}$ in the frequency range of the Ir t$_\text{2g}$ transitions for selected pressures together
with the Lorentz fit at 2.2\,GPa. (b) Corresponding optical conductivity spectra obtained via Kramers-Kronig analysis of the reflectivity
data and the Lorentz fit at 2.2\,GPa, together with the three main contributions A, B, and C.}\label{fig.d-d_pressure}
\end{figure}

The pressure-dependent reflectivity spectra $R_\text{s-d}$ in the frequency range of the Ir t\textsubscript{2g} excitations and the corresponding optical conductivity spectra are shown in Fig.~\ref{fig.d-d_pressure}(a) and (b), respectively.
At the lowest measured pressure (2.2\,GPa) one finds three main contributions to the optical conductivity, labeled {\bf A}, {\bf B}, and {\bf C}, whereby the contributions {\bf B} and {\bf C} are close in energy [see Fig.~\ref{fig.d-d_pressure}(b)].
According to theoretical calculations they correspond to excitations between the relativistic $j_{3/2}$ and $j_{1/2}$ orbitals~\cite{Kim.2014,Li.2015,Kim.2016c,Li.2017}: As discussed in our previous publication~\cite{Hermann.2017}, contribution \textbf{A} sums up the on-site $j_{3/2} \rightarrow j_{1/2}$ excitations, peak \textbf{B} is attributed to intersite $j_{1/2} \rightarrow j_{1/2}$ excitations, and peak \textbf{C} mainly involves intersite $j_{3/2} \rightarrow j_{1/2}$ excitations \cite{Kim.2014,Li.2015,Kim.2016c,Li.2017,Li.2017b}.
Additional contributions, namely intersite $j_{3/2}\rightarrow j_{1/2}$ and multiple spin-orbit (SO) excitons, were suggested for  peak B \cite{Li.2017b,Kim.2014}.
The spectral weight of contribution {\bf B} could be a measure of the indirect oxygen-mediated hopping and, thus, of the Kitaev interaction \cite{Li.2015}.
The fourth, weak contribution \textbf{D}, as observed for Na$_2$IrO$_3$ in the range 15500-17000\,cm$^{-1}$~\cite{Sohn.2013,Hermann.2017}, is even less pronounced in $\upalpha$-Li$_2$IrO$_3$ and therefore not included here.

\begin{figure}
	\includegraphics[width=0.46\textwidth]{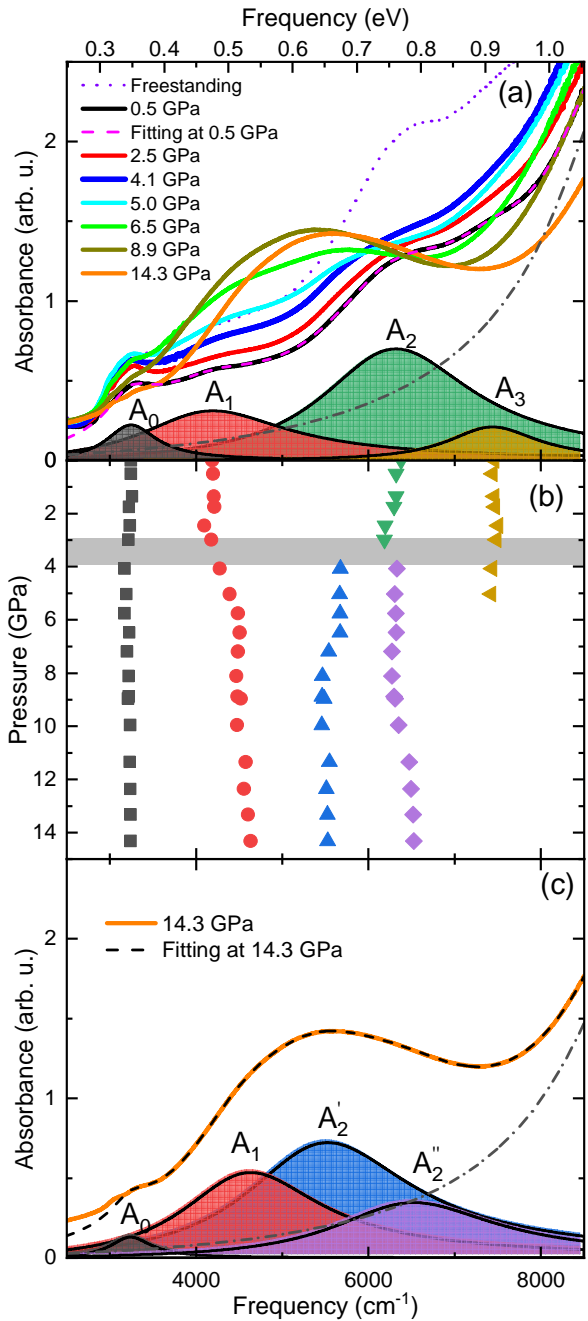}
	\caption{(a) Absorbance spectra of a thin crystal in the energy range around the on-site local transitions for selected pressures.
		For the absorbance spectrum at the lowest pressure (0.5 GPa) the Lorentz fitting together with the four distinct contributions $A_0$--$A_3$ is shown.
		The dimerization of the Ir--Ir network qualitatively affects the absorbance, as discussed in detail in the text. (b) Pressure dependence of the peak positions of the four contributions $A_0$--$A_3$. The grey shaded area marks the pressure where the sample dimerizes. Error bars from fitting are smaller than symbol size.
		(c) Absorbance spectrum at the highest studied pressure (14.3~GPa) together with the Lorentz fitting and the various contributions.
		The dash-dotted lines in (a) and (c) indicate higher-energy intersite $d$-$d$ excitations obtained by reflection measurements.}\label{fig.Absorbance}
\end{figure}

More detailed information on the very weak, low-energy $d$-$d$ transitions can be obtained from the pressure-dependent absorbance spectra measured on a 3~$\upmu$m thin sample (see Fig.~\ref{fig.Absorbance}).
At low pressures a good fit of the spectra is obtained with four contributions labelled \textbf{A\textsubscript{0}}--\textbf{A\textsubscript{3}}, which lie in the range of contribution \textbf{A} in the optical conductivity.
The comparison between the absorbance and optical conductivity spectra reveals that the main contribution to \textbf{A} is due to \textbf{A\textsubscript{2}}.
The energies of all contributions to the optical conductivity and absorbance at the lowest measured pressure, resp., are given in Table~\ref{tab.dd-low}.

\begin{table}[b]
	\caption{\label{tab.dd-low} Energetic position and assignment of the various observed contributions at the lowest pressure (2.2~GPa for \textbf{A}, \textbf{B}, \textbf{C} and 0.5~GPa for \textbf{A\textsubscript{0}}-\textbf{A\textsubscript{3}}) as explained in text.}
	\begin{ruledtabular}
		\begin{tabular}{ccrrr}
			& & \multicolumn{2}{c}{position} & assignment\\
			& & [eV] & [cm$^{-1}$] & \\\hline
			\rule{0pt}{2.6ex} & \textbf{A\textsubscript{0}} & 0.40 & 3250 & excitonic e-h excitations  \\
			& \textbf{A\textsubscript{1}} & 0.52 & 4200 & excitonic e-h excitations  \\
			\textbf{A} & & 0.79 & 6390 & on-site $j_{3/2}\rightarrow j_{1/2}$ \\
			& \textbf{A\textsubscript{2}} & 0.78 & 6320 & on-site $j_{3/2} \rightarrow j_{1/2}$ \\
			& \textbf{A\textsubscript{3}} & 0.92 & 7440 & on-site $j_{3/2} \rightarrow j_{1/2}$ \\
			\textbf{B} && 1.28 & 10350 & intersite $j_{1/2} \rightarrow j_{1/2}$ and \\
			&&&& intersite $j_{3/2} \rightarrow j_{1/2}$ and \\
			&&&& multiple SO excitons\\
			\textbf{C} && 1.64 & 13220 & intersite $j_{3/2} \rightarrow j_{1/2}$		
		\end{tabular}
	\end{ruledtabular}
\end{table}

\begin{figure}[t]
	\includegraphics[width=0.85\linewidth]{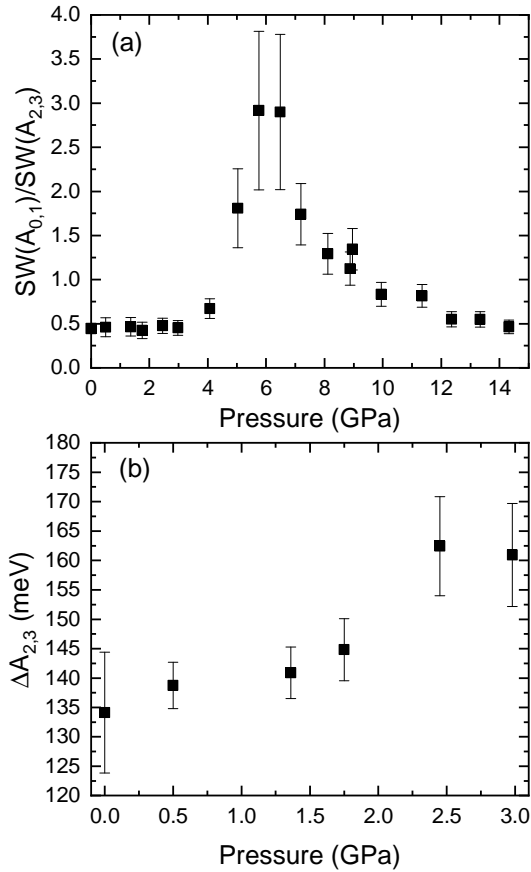}
	\caption{\label{fig.relA} (a) Relative spectral weight $SW(A_{0,1})/SW(A_{2,3})$ between the itinerant excitations \textbf{A\textsubscript{0}} and
\textbf{A\textsubscript{1}} and the local excitations \textbf{A\textsubscript{2}} and \textbf{A\textsubscript{3}}.
(b) Energy difference $\Delta$$A_{2,3}$ between the two local $j_{3/2} \rightarrow j_{1/2}$ excitations \textbf{A\textsubscript{2}} and \textbf{A\textsubscript{3}} before the dimerization at $P_c$.}
\end{figure}

For the interpretation of the observed features, we compare our optical spectra with the theoretical projected excitation spectrum of Na$_2$IrO$_3$ \cite{Kim.2014} and the theoretical and experimental RIXS spectra of $\upalpha$-Li$_2$IrO$_3$ and Na$_2$IrO$_3$ given in Refs.\ \cite{Kim.2014,Gretarsson.2013,Clancy.2018}, although different spectral weights are to be expected for our optical conductivity/absorbance spectra.
Accordingly, the energetic position of \textbf{A\textsubscript{2}} and \textbf{A\textsubscript{3}} fit well to the local $j_{3/2}\rightarrow j_{1/2}$ excitations \bibnote{We note that the contributions \textbf{A\textsubscript{2}} and \textbf{A\textsubscript{3}} would also fit energetically to interband transitions of QMOs, since $\upalpha$-Li$_2$IrO$_3$ is known to exhibit both Mott-insulating and QMO character~\cite{Kim.2016c,Hermann.2017} at ambient pressure.}.
The splitting of the local excitations into two contributions is due to the trigonal crystal field, which splits the $j_{3/2}$ level \cite{Gretarsson.2013}. Therefore, the energy difference $\Delta{A_{2,3}}$ between the contributions \textbf{A\textsubscript{2}} and \textbf{A\textsubscript{3}} may serve as a measure for the strength of the trigonal distortion of the IrO$_6$ octahedra \cite{Kim.2014}. The value of $\Delta{A_{2,3}}$ as a function of pressure
is depicted in Fig.\ \ref{fig.relA}(b).


The absorption features \textbf{A\textsubscript{0}} and \textbf{A\textsubscript{1}} can be attributed to excitonic electron-hole (e-h) excitations \cite{Kim.2014,Gretarsson.2013} and reflect the itinerant nature of the Ir 5$d$ orbitals \cite{Kim.2014}.
It is therefore intriguing to compare the spectral weight of the contributions \textbf{A\textsubscript{0}} and \textbf{A\textsubscript{1}} with that of the local excitations \textbf{A\textsubscript{2}} and \textbf{A\textsubscript{3}}.
To this end, the spectral weight $SW(A_{0,1})$ [$SW(A_{2,3})$] of the contributions \textbf{A\textsubscript{0}} and \textbf{A\textsubscript{1}} (\textbf{A\textsubscript{2}} and \textbf{A\textsubscript{3}}) was calculated from the area below the corresponding two Lorentz functions. The ratio of the spectral weights, $SW(A_{0,1})/SW(A_{2,3})$, could serve as a measure for the itinerant character of the material. The pressure dependence of the ratio $SW(A_{0,1})/SW(A_{2,3})$ is shown in Fig.\ \ref{fig.relA}(a).


Interestingly, the absorbance of $\upalpha$-Li$_2$IrO$_3$ at low pressures is very similar to the absorbance of Na$_2$IrO$_3$ at high pressures~\cite{Xi.2018}.
Note that in Ref.\ \cite{Xi.2018} the contribution $A_0$ appears only above $\approx$13.5~GPa.
(The fourth peak found in our present study is out of range for Ref.~\cite{Xi.2018}).
According to high-quality single crystal XRD data under pressure, $\upalpha$-Li$_2$IrO$_3$ is similar to Na$_2$IrO$_3$ at $\approx$29~GPa regarding lattice parameters~\cite{Hermann.2017}.
Electronically, these two materials are different due to different oxygen mediated hopping parameters, which depend not only on the Ir--O distances but also on Ir--O--Ir bond angles.
Nevertheless, the assignment of \textbf{A\textsubscript{0}} and \textbf{A\textsubscript{1}} in $\upalpha$-Li$_2$IrO$_3$ according to theoretical calculations for Na$_2$IrO$_3$ is justified, since no major energy shifts are found with pressure~\cite{Xi.2018,Hermann.2017} or by Li substitution~\cite{Hermann.2017} in this energy range.

\subsubsection{Pressure range \(P\)\(<\)\(P\textsubscript{c}\)}

In comparison to the ambient-pressure optical conductivity data, already at 2.2\,GPa the spectral weight of excitation \textbf{A}, which is mainly due to on-site excitations, is drastically reduced.
This drastic decrease was also observed by recent pressure-dependent RIXS experiments of Clancy \textit{et al.}~\cite{Clancy.2018}, where it was interpreted in
terms of an increasing QMO-nature. It was furthermore proposed that an itinerant QMO-like state develops for the pressure range 2~GPa$\leq$$P$$\leq$3~GPa, and that the relativistic $j\textsubscript{eff}$=1/2 ground state breaks down already at low pressure (\(P\)\(\leq\)2~GPa) \cite{Clancy.2018}.
In this low-pressure regime small distortions of the Ir--Ir network and the IrO$_6$ octahedra increase the intrahexagonal hopping parameters and suppress the interhexagon ones, favoring the QMO nature \cite{Foyevtsova.2013,Clancy.2018}.
Indeed, according to our absorbance data, the spectral weight of the on-site excitation \textbf{A\textsubscript{2}} at 0.5~GPa is nearly halved as compared to ambient pressure [see Fig.\ \ref{fig.Absorbance}(a)].
However, the relative spectral weight $SW(A_{0,1})/SW(A_{2,3})$ between the itinerant and local excitations does not change between 0.5 and 3.5~GPa [see Fig.\ \ref{fig.relA}(a)].
Therefore, the previously suggested \cite{Clancy.2018} increase in the itinerant character of $\alpha$-Li$_2$IrO$_3$ under pressure is not confirmed by our data
\bibnote{A weakening of the optically forbidden on-site excitations may in general also be due to a reduction of the trigonal distortion of the IrO$_6$ octahedra~\cite{Mazin.2012,Bhattacharjee.2012,Kim.2012}. However, according to the energy difference $\Delta{A_{2,3}}$ [see Fig.\ \ref{fig.relA}(b)] the opposite trend is observed for $\alpha$-Li$_2$IrO$_3$, namely an increase in the trigonal distortion with increasing pressure up to $P\textsubscript{c}$.}.
Nor have we confirmed the structural changes proposed in Ref. \cite{Clancy.2018}, as our single-crystal XRD data and phonon spectra both show only a smooth evolution of the structure below $P\textsubscript{c}$.

Besides the above-mentioned initial spectral weight reduction, we do not observe a further pressure-induced spectral weight decrease  for the low-energy ($\leq$1~eV) $d$-$d$ excitations  for pressures below $P_c$, contrary to the observations in RIXS studies~\cite{Clancy.2018}.
Instead, the spectral weight is slightly increased for all
four excitations \textbf{A\textsubscript{0}}-\textbf{A\textsubscript{3}} in a similar manner [see Figs.~\ref{fig.Absorbance}(a) and \ref{fig.relA}(a)].
The reason for the different behavior of the spectral weight in the RIXS and optical spectra might be related to different selection rules for the two techniques.
With increasing pressure, the local excitation \textbf{A\textsubscript{2}} shifts to lower energies, in agreement with RIXS results \cite{Clancy.2018}, whereas the positions of \textbf{A\textsubscript{0}}, \textbf{A\textsubscript{1}}, and the local excitation \textbf{A\textsubscript{3}} are independent of pressure within the error bar.
The shift of \textbf{A\textsubscript{2}} results in an increase in $\Delta{A_{2,3}}$ [see Fig.\ \ref{fig.relA}(b)], which suggests a pressure-induced increase in the trigonal distortion below the onset of dimerization~\cite{Kim.2014}.
The weak anomalies in some optical parameters already slightly below $P\textsubscript{c}$ (see Fig.\ \ref{fig.Absorbance}), suggest an enhancement of the QMO character due to the enhanced trigonal distortion \cite{Foyevtsova.2013}.

For the non-local transitions in the higher-energy range (excitations \textbf{B} and \textbf{C}) one observes only a slight pressure-induced decrease in the spectral weight below $P_\text{c}$\,=\,3.8\,GPa. The pressure-induced changes are further illustrated by the difference spectra $\Delta\sigma_1$ [see Fig.\ \ref{fig.dd_contour_diff}(b)] calculated according to $\Delta\sigma_1(\omega,P)$=$\sigma_1(\omega,P)$-$\sigma_1(\omega,2.2\,\text{GPa})$.

To summarize, below $P\textsubscript{c}$ no changes occur in the intersite physics, and there are only moderate changes in the on-site physics (energy difference of the on-site $j_{3/2} \rightarrow j_{1/2}$ transitions) caused by the change in the trigonal distortion.

\subsubsection{Pressure range {$P$}$>${$P_c$}}

Directly above $P\textsubscript{c}$, a pressure-induced increase in the spectral weight around the contribution \textbf{A} occurs in the optical conductivity (see Fig.\ \ref{fig.d-d_pressure}) and the absorbance [see Fig.\ \ref{fig.Absorbance}(a)]. According to the
pressure-dependent absorbance spectrum,
the main spectral weight increase is attributed to
\textbf{A\textsubscript{1}}, which concomitantly shifts to higher energies. The pressure-induced shift of \textbf{A\textsubscript{2}} to lower energies continues above $P_\text{c}$, but the spectrum can no longer be fitted well with only one Lorentz oscillator.
Instead, two Lorentz terms \textbf{A$_\textbf{2}^\mathbf{\prime}$} and \textbf{A$_\textbf{2}^\mathbf{\prime\prime}$} have to be included to obtain a good fit of the spectra [see Fig.\ \ref{fig.Absorbance}(a)].
We attribute the splitting to a proposed additional monoclinic distortion induced by Ir-Ir dimerization, which primarily affects the $j_{3/2}$ levels.
Concurrently, the spectral weight of excitation \textbf{A\textsubscript{3}}, which is connected to \textbf{A\textsubscript{2}} {\it via} the trigonal distortion below $P_c$, is decreased above $P\textsubscript{c}$, and the contribution \textbf{A\textsubscript{3}} is no longer observed for pressures above $\approx$5~GPa.

\begin{figure}[t]
	\includegraphics[width=\linewidth]{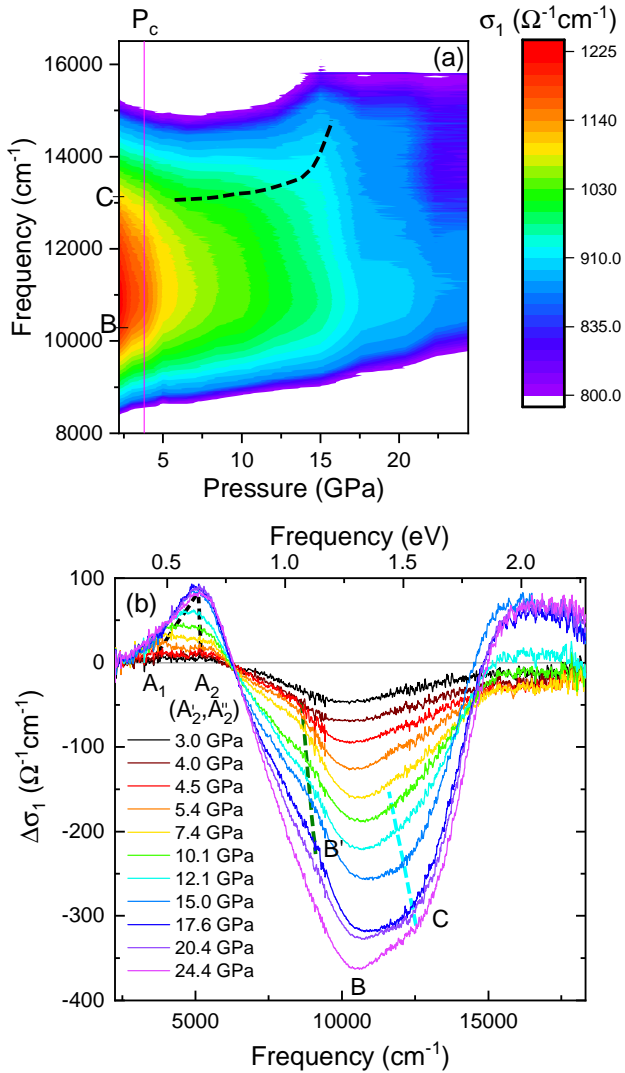}
	\caption{(a) Contour plot of the optical conductivity under external pressure. The critical pressure for the dimerization is indicated by the vertical line. The energy positions of the two strongest contributions B and C are indicated. The dashed line marks the shift of spectral weight at higher pressures. (b) Difference spectra $\Delta\sigma_1$ as defined in the text. The various dashed lines are discussed in the text.}\label{fig.dd_contour_diff}
\end{figure}

With further increasing the pressure up to around 8-10~GPa, \textbf{A$_\textbf{2}^\mathbf{\prime}$} shifts to lower frequencies so that \textbf{A$_\textbf{2}^\mathbf{\prime}$}, \textbf{A$_\textbf{2}^\mathbf{\prime\prime}$} and \textbf{A\textsubscript{1}} nearly combine to one broad peak, which can only be fitted well with three Lorentz oscillators.
Above $\sim$10~GPa this combined peak shifts to higher energies.
Interestingly, the spectral weight ratio $SW(A_{0,1})/SW(A_{2,3})$ shows a maximum at $P\textsubscript{c}$ before it decreases and reaches its low-pressure value [see Fig.\ \ref{fig.relA}(a)].
Whether the spectral weight ratio could still serve as a measure for the itinerant character of $\upalpha$-Li$_2$IrO$_3$ in its dimerized phase is, however, questionable.

Corresponding changes are revealed by the difference spectra $\Delta\sigma_1$ shown in Fig.\ \ref{fig.dd_contour_diff}(b):
Above {$\approx$}$P_\text{c}$ additional spectral weight grows around contribution \textbf{A}, in the frequency range 3000-6000\,cm$^{-1}$.
The increase in the spectral weight comprises two contributions, labeled \textbf{A\textsubscript{1}} and \textbf{A\textsubscript{2}}
\bibnote{According to the absorbance spectra [see Fig.\ \ref{fig.Absorbance}(c)] feature \textbf{A\textsubscript{2}} consists of two
contributions \textbf{A$_\textbf{2}^\mathbf{\prime}$} and \textbf{A$_\textbf{2}^\mathbf{\prime\prime}$}.}.
They shift towards each other and become nearly degenerate above $\sim$15\,GPa, where the increase in the spectral weight saturates.
The behavior of these two features with increasing pressure is indicated in Fig.~\ref{fig.dd_contour_diff}(b) by black, dashed lines.
The increasing spectral weight of the on-site excitations can be attributed to the increasing octahedral distortions ~\cite{Mazin.2012,Bhattacharjee.2012,Kim.2012}.

Also in the higher-energy range of the $d$-$d$ transitions significant pressure-induced changes occur well above $P_c$ according to the optical conductivity [see Fig.\ \ref{fig.d-d_pressure}(b)]: The trend of decreasing spectral weight is continued for contribution \textbf{B} and the spectral weight around contribution \textbf{C} shifts to higher frequencies with increasing pressure.
As seen more clearly in the contour plot [Fig.~\ref{fig.dd_contour_diff}(a)], this shift of the spectral weight is most pronounced for the pressure range 10-17\,GPa.

More detailed information on the changes can be obtained $via$ the difference spectra $\Delta\sigma_1$ depicted in Fig.\ \ref{fig.dd_contour_diff}(b):
i) From the lowest up to the highest measured pressure the spectral weight around \textbf{B} is decreased.
Since the contribution \textbf{B} is very broad, the whole measured frequency range is affected.
ii) A kink in the difference spectrum appears at {$\approx$}$P_\text{c}$, located at around 9000\,cm$^{-1}$, and is observable up to $\approx$15\,GPa [its pressure evolution is marked by a green, dashed line in Fig.~\ref{fig.dd_contour_diff}(b)].
It might be related to an additional feature \textbf{B$^\prime$} centered at around 9000\,cm$^{-1}$, whose origin could be the direct hopping between adjacent Ir$^{4+}$ ions~\cite{Li.2017}.
iii) An additional pressure-induced decrease in the spectral weight is observed in the range 11000-13000\,cm$^{-1}$, marked by the cyan dashed line, which is concomitant with the strong increase in the spectral weight for frequencies above 15000\,cm$^{-1}$.
These findings suggest a transfer of spectral weight from 11000-13000\,cm$^{-1}$ to the frequency range above 15000\,cm$^{-1}$, which is most pronounced for pressures above $\approx$8\,GPa. As a consequence, the optical conductivity has a clear three-peak profile at high pressures $\sim$20~GPa [see Fig.\ \ref{fig.d-d_pressure}(b)]. The additional spectral weight above 15000\,cm$^{-1}$ could be due to additional electronic bands appearing in the triclinic dimerized phase, as suggested by tight-binding band structure calculations for Na$_2$IrO$_3$~\cite{Hu.2018}.

For pressures {$P$}$>${$P\textsubscript{c}$} the large distortion of the honeycomb lattice results in the Ir--Ir dimerization and leads to a breakdown of the QMO-state \cite{Clancy.2018}.
The dimerized ground state of $\alpha$-Li$_2$IrO$_3$ above $P_c$ can neither be described by a relativistic $j\textsubscript{eff}$=1/2 state nor by an itinerant QMO-state. In Na$_2$IrO$_3$ in its dimerized phase, which is predicted at high pressure \cite{Hermann.2018,Hu.2018}, the dimerization introduces a Mott insulator to band insulator transition, whereby the low-energy excitations could possibly preserve their $j_{1/2}$-character~\cite{Hu.2018}.
Such a Mott to band insulator transition is common to other $j_{1/2}$ Mott insulators in their dimerized phase \cite{Takayama.2018,Antonov.2018,Bastien.2018}.
It is furthermore interesting to note that the pressure-induced changes in $\alpha$-Li$_2$IrO$_3$ seem to be quite similar to those in the three-dimensional analogue $\beta$-Li$_2$IrO$_3$, due to the similarities in the local environment of the Ir atoms: It was predicted that $\beta$-Li$_2$IrO$_3$ undergoes an electronic phase transition from Mott insulator to band insulators, concomitant to the structural phase transition to a dimerized phase \cite{Antonov.2018, Hu.2018,Takayama.2018}.
The three-peak profile of the optical conductivity of $\alpha$-Li$_2$IrO$_3$ at high pressures, reported here, appears to be indicative for the dimerized state in iridates~\cite{Takayama.2018}.

\section{Conclusion}
The pressure-induced dimerization of Ir--Ir bonds in $\upalpha$-Li$_2$IrO$_3$ at $P_\text{c}$=3.8~GPa leads to an anomaly in the hardening of the Li-based phonon modes.
The on-site $d$-$d$ transitions, located in the lower-energy range, are strongly affected by the dimerization, showing several anomalies.
In contrast, the intersite Ir $d$-$d$ transitions at higher energies are mainly affected for pressures well above $P_\text{c}$, showing a monotonic blueshift and decrease in spectral weight. At high pressure ($\sim$20~GPa) the optical conductivity consists of three main, well-separated contributions. This profile of the optical conductivity seems to be indicative for the dimerized state in iridates.

\begin{acknowledgments}
We thank Roser Valent{\'\i}, Steve Winter, and Ying Li for the fruitful discussions.
PG, AAT, and FF acknowledges financial support by the Deutsche Forschungsgemeinschaft (DFG) through TRR 80 and  SPP 1666. AJ acknowledges support from the DFG through Grant No. JE 748/1. AAT acknowledges financial support from the Federal Ministry for Education and Research via the Sofja-Kovalevskaya Award of Alexander von Humboldt Foundation.

\end{acknowledgments}


\begin{thebibliography}{56}%
\makeatletter
\providecommand \@ifxundefined [1]{%
 \@ifx{#1\undefined}
}%
\providecommand \@ifnum [1]{%
 \ifnum #1\expandafter \@firstoftwo
 \else \expandafter \@secondoftwo
 \fi
}%
\providecommand \@ifx [1]{%
 \ifx #1\expandafter \@firstoftwo
 \else \expandafter \@secondoftwo
 \fi
}%
\providecommand \natexlab [1]{#1}%
\providecommand \enquote  [1]{``#1''}%
\providecommand \bibnamefont  [1]{#1}%
\providecommand \bibfnamefont [1]{#1}%
\providecommand \citenamefont [1]{#1}%
\providecommand \href@noop [0]{\@secondoftwo}%
\providecommand \href [0]{\begingroup \@sanitize@url \@href}%
\providecommand \@href[1]{\@@startlink{#1}\@@href}%
\providecommand \@@href[1]{\endgroup#1\@@endlink}%
\providecommand \@sanitize@url [0]{\catcode `\\12\catcode `\$12\catcode
  `\&12\catcode `\#12\catcode `\^12\catcode `\_12\catcode `\%12\relax}%
\providecommand \@@startlink[1]{}%
\providecommand \@@endlink[0]{}%
\providecommand \url  [0]{\begingroup\@sanitize@url \@url }%
\providecommand \@url [1]{\endgroup\@href {#1}{\urlprefix }}%
\providecommand \urlprefix  [0]{URL }%
\providecommand \Eprint [0]{\href }%
\providecommand \doibase [0]{http://dx.doi.org/}%
\providecommand \selectlanguage [0]{\@gobble}%
\providecommand \bibinfo  [0]{\@secondoftwo}%
\providecommand \bibfield  [0]{\@secondoftwo}%
\providecommand \translation [1]{[#1]}%
\providecommand \BibitemOpen [0]{}%
\providecommand \bibitemStop [0]{}%
\providecommand \bibitemNoStop [0]{.\EOS\space}%
\providecommand \EOS [0]{\spacefactor3000\relax}%
\providecommand \BibitemShut  [1]{\csname bibitem#1\endcsname}%
\let\auto@bib@innerbib\@empty
\bibitem [{\citenamefont {Kitaev}(2006)}]{Kitaev.2006}%
  \BibitemOpen
  \bibfield  {author} {\bibinfo {author} {\bibfnamefont {A.}~\bibnamefont
  {Kitaev}},\ }\bibfield  {title} {\enquote {\bibinfo {title} {{Anyons in an
  exactly solved model and beyond}},}\ }\href {\doibase
  10.1016/j.aop.2005.10.005} {\bibfield  {journal} {\bibinfo  {journal} {{Ann.
  Phys. (NY)}}\ }\textbf {\bibinfo {volume} {321}},\ \bibinfo {pages} {2}
  (\bibinfo {year} {2006})}\BibitemShut {NoStop}%
\bibitem [{\citenamefont {Jackeli}\ and\ \citenamefont
  {Khaliullin}(2009)}]{Jackeli.2009}%
  \BibitemOpen
  \bibfield  {author} {\bibinfo {author} {\bibfnamefont {G.}~\bibnamefont
  {Jackeli}}\ and\ \bibinfo {author} {\bibfnamefont {G.}~\bibnamefont
  {Khaliullin}},\ }\bibfield  {title} {\enquote {\bibinfo {title} {{Mott
  Insulators in the Strong Spin-Orbit Coupling Limit: From Heisenberg to a
  Quantum Compass and Kitaev Models}},}\ }\href {\doibase
  10.1103/PhysRevLett.102.017205} {\bibfield  {journal} {\bibinfo  {journal}
  {{Phys. Rev. Lett.}}\ }\textbf {\bibinfo {volume} {102}},\ \bibinfo {pages}
  {017205} (\bibinfo {year} {2009})}\BibitemShut {NoStop}%
\bibitem [{\citenamefont {Chaloupka}\ \emph {et~al.}(2010)\citenamefont
  {Chaloupka}, \citenamefont {Jackeli},\ and\ \citenamefont
  {Khaliullin}}]{Chaloupka.2010}%
  \BibitemOpen
  \bibfield  {author} {\bibinfo {author} {\bibfnamefont {J.}~\bibnamefont
  {Chaloupka}}, \bibinfo {author} {\bibfnamefont {G.}~\bibnamefont {Jackeli}},
  \ and\ \bibinfo {author} {\bibfnamefont {G.}~\bibnamefont {Khaliullin}},\
  }\bibfield  {title} {\enquote {\bibinfo {title} {{Kitaev-Heisenberg Model on
  a Honeycomb Lattice: Possible Exotic Phases in Iridium Oxides
  $A_2$IrO$_3$}},}\ }\href {\doibase 10.1103/PhysRevLett.105.027204} {\bibfield
   {journal} {\bibinfo  {journal} {{Phys. Rev. Lett.}}\ }\textbf {\bibinfo
  {volume} {105}},\ \bibinfo {pages} {027204} (\bibinfo {year}
  {2010})}\BibitemShut {NoStop}%
\bibitem [{\citenamefont {Singh}\ \emph {et~al.}(2012)\citenamefont {Singh},
  \citenamefont {Manni}, \citenamefont {Reuther}, \citenamefont {Berlijn},
  \citenamefont {Thomale}, \citenamefont {Ku}, \citenamefont {Trebst},\ and\
  \citenamefont {Gegenwart}}]{Singh.2012}%
  \BibitemOpen
  \bibfield  {author} {\bibinfo {author} {\bibfnamefont {Y.}~\bibnamefont
  {Singh}}, \bibinfo {author} {\bibfnamefont {S.}~\bibnamefont {Manni}},
  \bibinfo {author} {\bibfnamefont {J.}~\bibnamefont {Reuther}}, \bibinfo
  {author} {\bibfnamefont {T.}~\bibnamefont {Berlijn}}, \bibinfo {author}
  {\bibfnamefont {R.}~\bibnamefont {Thomale}}, \bibinfo {author} {\bibfnamefont
  {W.}~\bibnamefont {Ku}}, \bibinfo {author} {\bibfnamefont {S.}~\bibnamefont
  {Trebst}}, \ and\ \bibinfo {author} {\bibfnamefont {P.}~\bibnamefont
  {Gegenwart}},\ }\bibfield  {title} {\enquote {\bibinfo {title} {{Relevance of
  the Heisenberg-Kitaev Model for the Honeycomb Lattice Iridates
  $A$$_2$IrO$_3$}},}\ }\href {\doibase 10.1103/PhysRevLett.108.127203}
  {\bibfield  {journal} {\bibinfo  {journal} {{Phys. Rev. Lett.}}\ }\textbf
  {\bibinfo {volume} {108}},\ \bibinfo {pages} {127203} (\bibinfo {year}
  {2012})}\BibitemShut {NoStop}%
\bibitem [{\citenamefont {Choi}\ \emph {et~al.}(2012)\citenamefont {Choi},
  \citenamefont {Coldea}, \citenamefont {Kolmogorov}, \citenamefont
  {Lancaster}, \citenamefont {Mazin}, \citenamefont {Blundell}, \citenamefont
  {Radaelli}, \citenamefont {Singh}, \citenamefont {Gegenwart}, \citenamefont
  {Choi}, \citenamefont {Cheong}, \citenamefont {Baker}, \citenamefont
  {Stock},\ and\ \citenamefont {Taylor}}]{Choi.2012}%
  \BibitemOpen
  \bibfield  {author} {\bibinfo {author} {\bibfnamefont {S.~K.}\ \bibnamefont
  {Choi}}, \bibinfo {author} {\bibfnamefont {R.}~\bibnamefont {Coldea}},
  \bibinfo {author} {\bibfnamefont {A.~N.}\ \bibnamefont {Kolmogorov}},
  \bibinfo {author} {\bibfnamefont {T.}~\bibnamefont {Lancaster}}, \bibinfo
  {author} {\bibfnamefont {I.~I.}\ \bibnamefont {Mazin}}, \bibinfo {author}
  {\bibfnamefont {S.~J.}\ \bibnamefont {Blundell}}, \bibinfo {author}
  {\bibfnamefont {P.~G.}\ \bibnamefont {Radaelli}}, \bibinfo {author}
  {\bibfnamefont {Yogesh}\ \bibnamefont {Singh}}, \bibinfo {author}
  {\bibfnamefont {P.}~\bibnamefont {Gegenwart}}, \bibinfo {author}
  {\bibfnamefont {K.~R.}\ \bibnamefont {Choi}}, \bibinfo {author}
  {\bibfnamefont {S.-W.}\ \bibnamefont {Cheong}}, \bibinfo {author}
  {\bibfnamefont {P.~J.}\ \bibnamefont {Baker}}, \bibinfo {author}
  {\bibfnamefont {C.}~\bibnamefont {Stock}}, \ and\ \bibinfo {author}
  {\bibfnamefont {J.}~\bibnamefont {Taylor}},\ }\bibfield  {title} {\enquote
  {\bibinfo {title} {{Spin Waves and Revised Crystal Structure of Honeycomb
  Iridate Na\textsubscript{2}IrO\textsubscript{3}}},}\ }\href {\doibase
  10.1103/PhysRevLett.108.127204} {\bibfield  {journal} {\bibinfo  {journal}
  {{Phys. Rev. Lett.}}\ }\textbf {\bibinfo {volume} {108}},\ \bibinfo {pages}
  {127204} (\bibinfo {year} {2012})}\BibitemShut {NoStop}%
\bibitem [{\citenamefont {Plumb}\ \emph {et~al.}(2014)\citenamefont {Plumb},
  \citenamefont {Clancy}, \citenamefont {Sandilands}, \citenamefont {Shankar},
  \citenamefont {Hu}, \citenamefont {Burch}, \citenamefont {Kee},\ and\
  \citenamefont {Kim}}]{Plumb.2014}%
  \BibitemOpen
  \bibfield  {author} {\bibinfo {author} {\bibfnamefont {K.~W.}\ \bibnamefont
  {Plumb}}, \bibinfo {author} {\bibfnamefont {J.~P.}\ \bibnamefont {Clancy}},
  \bibinfo {author} {\bibfnamefont {L.~J.}\ \bibnamefont {Sandilands}},
  \bibinfo {author} {\bibfnamefont {V.~Vijay}\ \bibnamefont {Shankar}},
  \bibinfo {author} {\bibfnamefont {Y.~F.}\ \bibnamefont {Hu}}, \bibinfo
  {author} {\bibfnamefont {K.~S.}\ \bibnamefont {Burch}}, \bibinfo {author}
  {\bibfnamefont {Hae-Young}\ \bibnamefont {Kee}}, \ and\ \bibinfo {author}
  {\bibfnamefont {Young-June}\ \bibnamefont {Kim}},\ }\bibfield  {title}
  {\enquote {\bibinfo {title} {{$\alpha$-RuCl$_3$: A spin-orbit assisted Mott
  insulator on a honeycomb lattice}},}\ }\href {\doibase
  10.1103/PhysRevB.90.041112} {\bibfield  {journal} {\bibinfo  {journal}
  {{Phys. Rev. B}}\ }\textbf {\bibinfo {volume} {90}},\ \bibinfo {pages}
  {041112} (\bibinfo {year} {2014})}\BibitemShut {NoStop}%
\bibitem [{\citenamefont {Chun}\ \emph {et~al.}(2015)\citenamefont {Chun},
  \citenamefont {Kim}, \citenamefont {Kim}, \citenamefont {Zheng},
  \citenamefont {Stoumpos}, \citenamefont {Malliakas}, \citenamefont
  {Mitchell}, \citenamefont {Mehlawat}, \citenamefont {Singh}, \citenamefont
  {Choi}, \citenamefont {Gog}, \citenamefont {Al-Zein}, \citenamefont {Sala},
  \citenamefont {Krisch}, \citenamefont {Chaloupka}, \citenamefont {Jackeli},
  \citenamefont {Khaliullin},\ and\ \citenamefont {Kim}}]{Chun.2015}%
  \BibitemOpen
  \bibfield  {author} {\bibinfo {author} {\bibfnamefont {S.~H.}\ \bibnamefont
  {Chun}}, \bibinfo {author} {\bibfnamefont {J.-W.}\ \bibnamefont {Kim}},
  \bibinfo {author} {\bibfnamefont {J.}~\bibnamefont {Kim}}, \bibinfo {author}
  {\bibfnamefont {H.}~\bibnamefont {Zheng}}, \bibinfo {author} {\bibfnamefont
  {C.~C.}\ \bibnamefont {Stoumpos}}, \bibinfo {author} {\bibfnamefont {C.~D.}\
  \bibnamefont {Malliakas}}, \bibinfo {author} {\bibfnamefont {J.~F.}\
  \bibnamefont {Mitchell}}, \bibinfo {author} {\bibfnamefont {K.}~\bibnamefont
  {Mehlawat}}, \bibinfo {author} {\bibfnamefont {Y.}~\bibnamefont {Singh}},
  \bibinfo {author} {\bibfnamefont {Y.}~\bibnamefont {Choi}}, \bibinfo {author}
  {\bibfnamefont {T.}~\bibnamefont {Gog}}, \bibinfo {author} {\bibfnamefont
  {A.}~\bibnamefont {Al-Zein}}, \bibinfo {author} {\bibfnamefont {M.~Moretti}\
  \bibnamefont {Sala}}, \bibinfo {author} {\bibfnamefont {M.}~\bibnamefont
  {Krisch}}, \bibinfo {author} {\bibfnamefont {J.}~\bibnamefont {Chaloupka}},
  \bibinfo {author} {\bibfnamefont {G.}~\bibnamefont {Jackeli}}, \bibinfo
  {author} {\bibfnamefont {G.}~\bibnamefont {Khaliullin}}, \ and\ \bibinfo
  {author} {\bibfnamefont {B.~J.}\ \bibnamefont {Kim}},\ }\bibfield  {title}
  {\enquote {\bibinfo {title} {{Direct evidence for dominant bond-directional
  interactions in a honeycomb lattice iridate Na$_2$IrO$_3$}},}\ }\href
  {\doibase 10.1038/nphys3322} {\bibfield  {journal} {\bibinfo  {journal}
  {{Nat. Phys.}}\ }\textbf {\bibinfo {volume} {11}},\ \bibinfo {pages} {462}
  (\bibinfo {year} {2015})}\BibitemShut {NoStop}%
\bibitem [{\citenamefont {Winter}\ \emph {et~al.}(2017)\citenamefont {Winter},
  \citenamefont {Tsirlin}, \citenamefont {Daghofer}, \citenamefont {van~den
  Brink}, \citenamefont {Singh}, \citenamefont {Gegenwart},\ and\ \citenamefont
  {Valent\'\i}}]{Winter.2017}%
  \BibitemOpen
  \bibfield  {author} {\bibinfo {author} {\bibfnamefont {S.~M.}\ \bibnamefont
  {Winter}}, \bibinfo {author} {\bibfnamefont {A.~A.}\ \bibnamefont {Tsirlin}},
  \bibinfo {author} {\bibfnamefont {M.}~\bibnamefont {Daghofer}}, \bibinfo
  {author} {\bibfnamefont {J.}~\bibnamefont {van~den Brink}}, \bibinfo {author}
  {\bibfnamefont {Y.}~\bibnamefont {Singh}}, \bibinfo {author} {\bibfnamefont
  {P.}~\bibnamefont {Gegenwart}}, \ and\ \bibinfo {author} {\bibfnamefont
  {R.}~\bibnamefont {Valent\'\i}},\ }\bibfield  {title} {\enquote {\bibinfo
  {title} {{Models and materials for generalized Kitaev magnetism}},}\ }\href
  {\doibase 10.1088/1361-648X/aa8cf5} {\bibfield  {journal} {\bibinfo
  {journal} {{J. Phys.: Condens. Matter}}\ }\textbf {\bibinfo {volume} {29}},\
  \bibinfo {pages} {493002} (\bibinfo {year} {2017})}\BibitemShut {NoStop}%
\bibitem [{\citenamefont {Sears}\ \emph {et~al.}(2015)\citenamefont {Sears},
  \citenamefont {Songvilay}, \citenamefont {Plumb}, \citenamefont {Clancy},
  \citenamefont {Qiu}, \citenamefont {Zhao}, \citenamefont {Parshall},\ and\
  \citenamefont {Kim}}]{Sears.2015}%
  \BibitemOpen
  \bibfield  {author} {\bibinfo {author} {\bibfnamefont {J.~A.}\ \bibnamefont
  {Sears}}, \bibinfo {author} {\bibfnamefont {M.}~\bibnamefont {Songvilay}},
  \bibinfo {author} {\bibfnamefont {K.~W.}\ \bibnamefont {Plumb}}, \bibinfo
  {author} {\bibfnamefont {J.~P.}\ \bibnamefont {Clancy}}, \bibinfo {author}
  {\bibfnamefont {Y.}~\bibnamefont {Qiu}}, \bibinfo {author} {\bibfnamefont
  {Y.}~\bibnamefont {Zhao}}, \bibinfo {author} {\bibfnamefont {D.}~\bibnamefont
  {Parshall}}, \ and\ \bibinfo {author} {\bibfnamefont {Young-June}\
  \bibnamefont {Kim}},\ }\bibfield  {title} {\enquote {\bibinfo {title}
  {{Magnetic order in $\alpha$--RuCl$_3$: A honeycomb-lattice quantum magnet
  with strong spin-orbit coupling}},}\ }\href {\doibase
  10.1103/PhysRevB.91.144420} {\bibfield  {journal} {\bibinfo  {journal}
  {{Phys. Rev. B}}\ }\textbf {\bibinfo {volume} {91}},\ \bibinfo {pages}
  {144420} (\bibinfo {year} {2015})}\BibitemShut {NoStop}%
\bibitem [{\citenamefont {Banerjee}\ \emph {et~al.}(2016)\citenamefont
  {Banerjee}, \citenamefont {Bridges}, \citenamefont {Yan}, \citenamefont
  {Aczel}, \citenamefont {Li}, \citenamefont {Stone}, \citenamefont {Granroth},
  \citenamefont {Lumsden}, \citenamefont {Yiu}, \citenamefont {Knolle},
  \citenamefont {Bhattacharjee}, \citenamefont {Kovrizhin}, \citenamefont
  {Moessner}, \citenamefont {Tennant}, \citenamefont {Mandrus},\ and\
  \citenamefont {Nagler}}]{Banerjee.2016}%
  \BibitemOpen
  \bibfield  {author} {\bibinfo {author} {\bibfnamefont {A.}~\bibnamefont
  {Banerjee}}, \bibinfo {author} {\bibfnamefont {C.~A.}\ \bibnamefont
  {Bridges}}, \bibinfo {author} {\bibfnamefont {J-Q}\ \bibnamefont {Yan}},
  \bibinfo {author} {\bibfnamefont {A.~A.}\ \bibnamefont {Aczel}}, \bibinfo
  {author} {\bibfnamefont {L.}~\bibnamefont {Li}}, \bibinfo {author}
  {\bibfnamefont {M.~B.}\ \bibnamefont {Stone}}, \bibinfo {author}
  {\bibfnamefont {G.~E.}\ \bibnamefont {Granroth}}, \bibinfo {author}
  {\bibfnamefont {M.~D.}\ \bibnamefont {Lumsden}}, \bibinfo {author}
  {\bibfnamefont {Y.}~\bibnamefont {Yiu}}, \bibinfo {author} {\bibfnamefont
  {J.}~\bibnamefont {Knolle}}, \bibinfo {author} {\bibfnamefont
  {S.}~\bibnamefont {Bhattacharjee}}, \bibinfo {author} {\bibfnamefont {D.~L.}\
  \bibnamefont {Kovrizhin}}, \bibinfo {author} {\bibfnamefont {R.}~\bibnamefont
  {Moessner}}, \bibinfo {author} {\bibfnamefont {D.~A.}\ \bibnamefont
  {Tennant}}, \bibinfo {author} {\bibfnamefont {D.~G.}\ \bibnamefont
  {Mandrus}}, \ and\ \bibinfo {author} {\bibfnamefont {S.~E.}\ \bibnamefont
  {Nagler}},\ }\bibfield  {title} {\enquote {\bibinfo {title} {{Proximate
  Kitaev quantum spin liquid behaviour in a honeycomb magnet}},}\ }\href
  {\doibase 10.1038/nmat4604} {\bibfield  {journal} {\bibinfo  {journal} {{Nat.
  Mater.}}\ }\textbf {\bibinfo {volume} {15}},\ \bibinfo {pages} {733}
  (\bibinfo {year} {2016})}\BibitemShut {NoStop}%
\bibitem [{\citenamefont {Williams}\ \emph {et~al.}(2016)\citenamefont
  {Williams}, \citenamefont {Johnson}, \citenamefont {Freund}, \citenamefont
  {Choi}, \citenamefont {Jesche}, \citenamefont {Kimchi}, \citenamefont
  {Manni}, \citenamefont {Bombardi}, \citenamefont {Manuel}, \citenamefont
  {Gegenwart},\ and\ \citenamefont {Coldea}}]{Williams.2016}%
  \BibitemOpen
  \bibfield  {author} {\bibinfo {author} {\bibfnamefont {S.~C.}\ \bibnamefont
  {Williams}}, \bibinfo {author} {\bibfnamefont {R.~D.}\ \bibnamefont
  {Johnson}}, \bibinfo {author} {\bibfnamefont {F.}~\bibnamefont {Freund}},
  \bibinfo {author} {\bibfnamefont {Sungkyun}\ \bibnamefont {Choi}}, \bibinfo
  {author} {\bibfnamefont {A.}~\bibnamefont {Jesche}}, \bibinfo {author}
  {\bibfnamefont {I.}~\bibnamefont {Kimchi}}, \bibinfo {author} {\bibfnamefont
  {S.}~\bibnamefont {Manni}}, \bibinfo {author} {\bibfnamefont
  {A.}~\bibnamefont {Bombardi}}, \bibinfo {author} {\bibfnamefont
  {P.}~\bibnamefont {Manuel}}, \bibinfo {author} {\bibfnamefont
  {P.}~\bibnamefont {Gegenwart}}, \ and\ \bibinfo {author} {\bibfnamefont
  {R.}~\bibnamefont {Coldea}},\ }\bibfield  {title} {\enquote {\bibinfo {title}
  {{Incommensurate counterrotating magnetic order stabilized by Kitaev
  interactions in the layered honeycomb
  $\alpha$-Li\textsubscript{2}IrO\textsubscript{3}}},}\ }\href {\doibase
  10.1103/PhysRevB.93.195158} {\bibfield  {journal} {\bibinfo  {journal}
  {{Phys. Rev. B}}\ }\textbf {\bibinfo {volume} {93}},\ \bibinfo {pages}
  {195158} (\bibinfo {year} {2016})}\BibitemShut {NoStop}%
\bibitem [{\citenamefont {Wolter}\ \emph {et~al.}(2017)\citenamefont {Wolter},
  \citenamefont {Corredor}, \citenamefont {Janssen}, \citenamefont {Nenkov},
  \citenamefont {Sch\"onecker}, \citenamefont {Do}, \citenamefont {Choi},
  \citenamefont {Albrecht}, \citenamefont {Hunger}, \citenamefont {Doert},
  \citenamefont {Vojta},\ and\ \citenamefont {B\"uchner}}]{Wolter.2017}%
  \BibitemOpen
  \bibfield  {author} {\bibinfo {author} {\bibfnamefont {A.~U.~B.}\
  \bibnamefont {Wolter}}, \bibinfo {author} {\bibfnamefont {L.~T.}\
  \bibnamefont {Corredor}}, \bibinfo {author} {\bibfnamefont {L.}~\bibnamefont
  {Janssen}}, \bibinfo {author} {\bibfnamefont {K.}~\bibnamefont {Nenkov}},
  \bibinfo {author} {\bibfnamefont {S.}~\bibnamefont {Sch\"onecker}}, \bibinfo
  {author} {\bibfnamefont {S.-H.}\ \bibnamefont {Do}}, \bibinfo {author}
  {\bibfnamefont {K.-Y.}\ \bibnamefont {Choi}}, \bibinfo {author}
  {\bibfnamefont {R.}~\bibnamefont {Albrecht}}, \bibinfo {author}
  {\bibfnamefont {J.}~\bibnamefont {Hunger}}, \bibinfo {author} {\bibfnamefont
  {T.}~\bibnamefont {Doert}}, \bibinfo {author} {\bibfnamefont
  {M.}~\bibnamefont {Vojta}}, \ and\ \bibinfo {author} {\bibfnamefont
  {B.}~\bibnamefont {B\"uchner}},\ }\bibfield  {title} {\enquote {\bibinfo
  {title} {{Field-induced quantum criticality in the Kitaev system
  $\alpha-\mathrm{RuCl}_{3}$}},}\ }\href {\doibase 10.1103/PhysRevB.96.041405}
  {\bibfield  {journal} {\bibinfo  {journal} {Phys. Rev. B}\ }\textbf {\bibinfo
  {volume} {96}},\ \bibinfo {pages} {041405} (\bibinfo {year}
  {2017})}\BibitemShut {NoStop}%
\bibitem [{\citenamefont {Baek}\ \emph {et~al.}(2017)\citenamefont {Baek},
  \citenamefont {Do}, \citenamefont {Choi}, \citenamefont {Kwon}, \citenamefont
  {Wolter}, \citenamefont {Nishimoto}, \citenamefont {van~den Brink},\ and\
  \citenamefont {B\"uchner}}]{Baek.2017}%
  \BibitemOpen
  \bibfield  {author} {\bibinfo {author} {\bibfnamefont {S.-H.}\ \bibnamefont
  {Baek}}, \bibinfo {author} {\bibfnamefont {S.-H.}\ \bibnamefont {Do}},
  \bibinfo {author} {\bibfnamefont {K.-Y.}\ \bibnamefont {Choi}}, \bibinfo
  {author} {\bibfnamefont {Y.~S.}\ \bibnamefont {Kwon}}, \bibinfo {author}
  {\bibfnamefont {A.~U.~B.}\ \bibnamefont {Wolter}}, \bibinfo {author}
  {\bibfnamefont {S.}~\bibnamefont {Nishimoto}}, \bibinfo {author}
  {\bibfnamefont {J.}~\bibnamefont {van~den Brink}}, \ and\ \bibinfo {author}
  {\bibfnamefont {B.}~\bibnamefont {B\"uchner}},\ }\bibfield  {title} {\enquote
  {\bibinfo {title} {{Evidence for a Field-Induced Quantum Spin Liquid in
  $\alpha$-${\mathrm{RuCl}}_{3}$}},}\ }\href {\doibase
  10.1103/PhysRevLett.119.037201} {\bibfield  {journal} {\bibinfo  {journal}
  {Phys. Rev. Lett.}\ }\textbf {\bibinfo {volume} {119}},\ \bibinfo {pages}
  {037201} (\bibinfo {year} {2017})}\BibitemShut {NoStop}%
\bibitem [{\citenamefont {Janssen}\ \emph {et~al.}(2016)\citenamefont
  {Janssen}, \citenamefont {Andrade},\ and\ \citenamefont
  {Vojta}}]{Janssen.2016}%
  \BibitemOpen
  \bibfield  {author} {\bibinfo {author} {\bibfnamefont {L.}~\bibnamefont
  {Janssen}}, \bibinfo {author} {\bibfnamefont {E.~C.}\ \bibnamefont
  {Andrade}}, \ and\ \bibinfo {author} {\bibfnamefont {M.}~\bibnamefont
  {Vojta}},\ }\bibfield  {title} {\enquote {\bibinfo {title}
  {{Honeycomb-Lattice Heisenberg-Kitaev Model in a Magnetic Field: Spin
  Canting, Metamagnetism, and Vortex Crystals}},}\ }\href {\doibase
  10.1103/PhysRevLett.117.277202} {\bibfield  {journal} {\bibinfo  {journal}
  {Phys. Rev. Lett.}\ }\textbf {\bibinfo {volume} {117}},\ \bibinfo {pages}
  {277202} (\bibinfo {year} {2016})}\BibitemShut {NoStop}%
\bibitem [{\citenamefont {Winter}\ \emph {et~al.}(2018)\citenamefont {Winter},
  \citenamefont {Riedl}, \citenamefont {Kaib}, \citenamefont {Coldea},\ and\
  \citenamefont {Valent\'{\i}}}]{Winter.2018}%
  \BibitemOpen
  \bibfield  {author} {\bibinfo {author} {\bibfnamefont {S.~M.}\ \bibnamefont
  {Winter}}, \bibinfo {author} {\bibfnamefont {K.}~\bibnamefont {Riedl}},
  \bibinfo {author} {\bibfnamefont {D.}~\bibnamefont {Kaib}}, \bibinfo {author}
  {\bibfnamefont {R.}~\bibnamefont {Coldea}}, \ and\ \bibinfo {author}
  {\bibfnamefont {R.}~\bibnamefont {Valent\'{\i}}},\ }\bibfield  {title}
  {\enquote {\bibinfo {title} {Probing $\alpha$-rucl$_3$ beyond magnetic order:
  Effects of temperature and magnetic field},}\ }\href {\doibase
  10.1103/PhysRevLett.120.077203} {\bibfield  {journal} {\bibinfo  {journal}
  {Phys. Rev. Lett.}\ }\textbf {\bibinfo {volume} {120}},\ \bibinfo {pages}
  {077203} (\bibinfo {year} {2018})}\BibitemShut {NoStop}%
\bibitem [{\citenamefont {Takagi}\ \emph {et~al.}()\citenamefont {Takagi},
  \citenamefont {Takayama}, \citenamefont {Jackeli}, \citenamefont
  {Khaliullin},\ and\ \citenamefont {Nagler}}]{Takagi.2019}%
  \BibitemOpen
  \bibfield  {author} {\bibinfo {author} {\bibfnamefont {H.}~\bibnamefont
  {Takagi}}, \bibinfo {author} {\bibfnamefont {T.}~\bibnamefont {Takayama}},
  \bibinfo {author} {\bibfnamefont {G.}~\bibnamefont {Jackeli}}, \bibinfo
  {author} {\bibfnamefont {G.}~\bibnamefont {Khaliullin}}, \ and\ \bibinfo
  {author} {\bibfnamefont {S.~E.}\ \bibnamefont {Nagler}},\ }\bibfield  {title}
  {\enquote {\bibinfo {title} {{Kitaev quantum spin liquid - concept and
  materialization}},}\ }\href@noop {} {\bibinfo  {journal} {arXiv:1903.08081}\
  }\BibitemShut {NoStop}%
\bibitem [{\citenamefont {Hermann}\ \emph {et~al.}(2017)\citenamefont
  {Hermann}, \citenamefont {Ebad-Allah}, \citenamefont {Freund}, \citenamefont
  {Pietsch}, \citenamefont {Jesche}, \citenamefont {Tsirlin}, \citenamefont
  {Deisenhofer}, \citenamefont {Hanfland}, \citenamefont {Gegenwart},\ and\
  \citenamefont {Kuntscher}}]{Hermann.2017}%
  \BibitemOpen
\bibfield  {journal} {  }\bibfield  {author} {\bibinfo {author} {\bibfnamefont
  {V.}~\bibnamefont {Hermann}}, \bibinfo {author} {\bibfnamefont
  {J.}~\bibnamefont {Ebad-Allah}}, \bibinfo {author} {\bibfnamefont
  {F.}~\bibnamefont {Freund}}, \bibinfo {author} {\bibfnamefont {I.~M.}\
  \bibnamefont {Pietsch}}, \bibinfo {author} {\bibfnamefont {A.}~\bibnamefont
  {Jesche}}, \bibinfo {author} {\bibfnamefont {A.~A.}\ \bibnamefont {Tsirlin}},
  \bibinfo {author} {\bibfnamefont {J.}~\bibnamefont {Deisenhofer}}, \bibinfo
  {author} {\bibfnamefont {M.}~\bibnamefont {Hanfland}}, \bibinfo {author}
  {\bibfnamefont {P.}~\bibnamefont {Gegenwart}}, \ and\ \bibinfo {author}
  {\bibfnamefont {C.~A.}\ \bibnamefont {Kuntscher}},\ }\bibfield  {title}
  {\enquote {\bibinfo {title} {{High-pressure versus isoelectronic doping
  effect on the honeycomb iridate Na\textsubscript{2}IrO\textsubscript{3}}},}\
  }\href {\doibase 10.1103/PhysRevB.96.195137} {\bibfield  {journal} {\bibinfo
  {journal} {{Phys. Rev. B}}\ }\textbf {\bibinfo {volume} {96}},\ \bibinfo
  {pages} {195137} (\bibinfo {year} {2017})}\BibitemShut {NoStop}%
\bibitem [{\citenamefont {Biesner}\ \emph {et~al.}(2018)\citenamefont
  {Biesner}, \citenamefont {Biswas}, \citenamefont {Li}, \citenamefont {Saito},
  \citenamefont {Pustogow}, \citenamefont {Altmeyer}, \citenamefont {Wolter},
  \citenamefont {B\"uchner}, \citenamefont {Roslova}, \citenamefont {Doert},
  \citenamefont {Winter}, \citenamefont {Valent\'{\i}},\ and\ \citenamefont
  {Dressel}}]{Biesner.2018}%
  \BibitemOpen
  \bibfield  {author} {\bibinfo {author} {\bibfnamefont {T.}~\bibnamefont
  {Biesner}}, \bibinfo {author} {\bibfnamefont {S.}~\bibnamefont {Biswas}},
  \bibinfo {author} {\bibfnamefont {W.}~\bibnamefont {Li}}, \bibinfo {author}
  {\bibfnamefont {Y.}~\bibnamefont {Saito}}, \bibinfo {author} {\bibfnamefont
  {A.}~\bibnamefont {Pustogow}}, \bibinfo {author} {\bibfnamefont
  {M.}~\bibnamefont {Altmeyer}}, \bibinfo {author} {\bibfnamefont {A.~U.~B.}\
  \bibnamefont {Wolter}}, \bibinfo {author} {\bibfnamefont {B.}~\bibnamefont
  {B\"uchner}}, \bibinfo {author} {\bibfnamefont {M.}~\bibnamefont {Roslova}},
  \bibinfo {author} {\bibfnamefont {T.}~\bibnamefont {Doert}}, \bibinfo
  {author} {\bibfnamefont {S.~M.}\ \bibnamefont {Winter}}, \bibinfo {author}
  {\bibfnamefont {R.}~\bibnamefont {Valent\'{\i}}}, \ and\ \bibinfo {author}
  {\bibfnamefont {M.}~\bibnamefont {Dressel}},\ }\bibfield  {title} {\enquote
  {\bibinfo {title} {{Detuning the honeycomb of
  $\ensuremath{\alpha}\text{\ensuremath{-}}{\mathrm{RuCl}}_{3}$:
  Pressure-dependent optical studies reveal broken symmetry}},}\ }\href
  {\doibase 10.1103/PhysRevB.97.220401} {\bibfield  {journal} {\bibinfo
  {journal} {Phys. Rev. B}\ }\textbf {\bibinfo {volume} {97}},\ \bibinfo
  {pages} {220401} (\bibinfo {year} {2018})}\BibitemShut {NoStop}%
\bibitem [{\citenamefont {Simutis}\ \emph {et~al.}(2018)\citenamefont
  {Simutis}, \citenamefont {Barbero}, \citenamefont {Rolfs}, \citenamefont
  {Leroy-Calatayud}, \citenamefont {Mehlawat}, \citenamefont {Khasanov},
  \citenamefont {Luetkens}, \citenamefont {Pomjakushina}, \citenamefont
  {Singh}, \citenamefont {Ott}, \citenamefont {Mesot}, \citenamefont {Amato},\
  and\ \citenamefont {Shiroka}}]{Simutis.2018}%
  \BibitemOpen
  \bibfield  {author} {\bibinfo {author} {\bibfnamefont {G.}~\bibnamefont
  {Simutis}}, \bibinfo {author} {\bibfnamefont {N.}~\bibnamefont {Barbero}},
  \bibinfo {author} {\bibfnamefont {K.}~\bibnamefont {Rolfs}}, \bibinfo
  {author} {\bibfnamefont {P.}~\bibnamefont {Leroy-Calatayud}}, \bibinfo
  {author} {\bibfnamefont {K.}~\bibnamefont {Mehlawat}}, \bibinfo {author}
  {\bibfnamefont {R.}~\bibnamefont {Khasanov}}, \bibinfo {author}
  {\bibfnamefont {H.}~\bibnamefont {Luetkens}}, \bibinfo {author}
  {\bibfnamefont {E.}~\bibnamefont {Pomjakushina}}, \bibinfo {author}
  {\bibfnamefont {Y.}~\bibnamefont {Singh}}, \bibinfo {author} {\bibfnamefont
  {H.-R.}\ \bibnamefont {Ott}}, \bibinfo {author} {\bibfnamefont
  {J.}~\bibnamefont {Mesot}}, \bibinfo {author} {\bibfnamefont
  {A.}~\bibnamefont {Amato}}, \ and\ \bibinfo {author} {\bibfnamefont
  {T.}~\bibnamefont {Shiroka}},\ }\bibfield  {title} {\enquote {\bibinfo
  {title} {{Chemical and hydrostatic-pressure effects on the Kitaev honeycomb
  material ${\mathrm{Na}}_{2}{\mathrm{IrO}}_{3}$}},}\ }\href {\doibase
  10.1103/PhysRevB.98.104421} {\bibfield  {journal} {\bibinfo  {journal} {Phys.
  Rev. B}\ }\textbf {\bibinfo {volume} {98}},\ \bibinfo {pages} {104421}
  (\bibinfo {year} {2018})}\BibitemShut {NoStop}%
\bibitem [{\citenamefont {Wang}\ \emph {et~al.}(2018)\citenamefont {Wang},
  \citenamefont {Guo}, \citenamefont {Tafti}, \citenamefont {Hegg},
  \citenamefont {Sen}, \citenamefont {Sidorov}, \citenamefont {Wang},
  \citenamefont {Cai}, \citenamefont {Yi}, \citenamefont {Zhou}, \citenamefont
  {Wang}, \citenamefont {Zhang}, \citenamefont {Yang}, \citenamefont {Li},
  \citenamefont {Li}, \citenamefont {Li}, \citenamefont {Liu}, \citenamefont
  {Shi}, \citenamefont {Ku}, \citenamefont {Wu}, \citenamefont {Cava},\ and\
  \citenamefont {Sun}}]{Wang.2018}%
  \BibitemOpen
  \bibfield  {author} {\bibinfo {author} {\bibfnamefont {Z.}~\bibnamefont
  {Wang}}, \bibinfo {author} {\bibfnamefont {J.}~\bibnamefont {Guo}}, \bibinfo
  {author} {\bibfnamefont {F.~F.}\ \bibnamefont {Tafti}}, \bibinfo {author}
  {\bibfnamefont {A.}~\bibnamefont {Hegg}}, \bibinfo {author} {\bibfnamefont
  {S.}~\bibnamefont {Sen}}, \bibinfo {author} {\bibfnamefont {V.~A.}\
  \bibnamefont {Sidorov}}, \bibinfo {author} {\bibfnamefont {L.}~\bibnamefont
  {Wang}}, \bibinfo {author} {\bibfnamefont {S.}~\bibnamefont {Cai}}, \bibinfo
  {author} {\bibfnamefont {W.}~\bibnamefont {Yi}}, \bibinfo {author}
  {\bibfnamefont {Y.}~\bibnamefont {Zhou}}, \bibinfo {author} {\bibfnamefont
  {H.}~\bibnamefont {Wang}}, \bibinfo {author} {\bibfnamefont {S.}~\bibnamefont
  {Zhang}}, \bibinfo {author} {\bibfnamefont {K.}~\bibnamefont {Yang}},
  \bibinfo {author} {\bibfnamefont {A.}~\bibnamefont {Li}}, \bibinfo {author}
  {\bibfnamefont {X.}~\bibnamefont {Li}}, \bibinfo {author} {\bibfnamefont
  {Y.}~\bibnamefont {Li}}, \bibinfo {author} {\bibfnamefont {J.}~\bibnamefont
  {Liu}}, \bibinfo {author} {\bibfnamefont {Y.}~\bibnamefont {Shi}}, \bibinfo
  {author} {\bibfnamefont {W.}~\bibnamefont {Ku}}, \bibinfo {author}
  {\bibfnamefont {Q.}~\bibnamefont {Wu}}, \bibinfo {author} {\bibfnamefont
  {.~J.}\ \bibnamefont {Cava}}, \ and\ \bibinfo {author} {\bibfnamefont
  {L.}~\bibnamefont {Sun}},\ }\bibfield  {title} {\enquote {\bibinfo {title}
  {{Pressure-induced melting of magnetic order and emergence of a new quantum
  state in
  $\ensuremath{\alpha}\text{\ensuremath{-}}\mathrm{RuC}{\mathrm{l}}_{3}$}},}\
  }\href {\doibase 10.1103/PhysRevB.97.245149} {\bibfield  {journal} {\bibinfo
  {journal} {Phys. Rev. B}\ }\textbf {\bibinfo {volume} {97}},\ \bibinfo
  {pages} {245149} (\bibinfo {year} {2018})}\BibitemShut {NoStop}%
\bibitem [{\citenamefont {Hermann}\ \emph {et~al.}(2018)\citenamefont
  {Hermann}, \citenamefont {Altmeyer}, \citenamefont {Ebad-Allah},
  \citenamefont {Freund}, \citenamefont {Jesche}, \citenamefont {Tsirlin},
  \citenamefont {Hanfland}, \citenamefont {Gegenwart}, \citenamefont {Mazin},
  \citenamefont {Khomskii}, \citenamefont {Valent\'\i},\ and\ \citenamefont
  {Kuntscher}}]{Hermann.2018}%
  \BibitemOpen
  \bibfield  {author} {\bibinfo {author} {\bibfnamefont {V.}~\bibnamefont
  {Hermann}}, \bibinfo {author} {\bibfnamefont {M.}~\bibnamefont {Altmeyer}},
  \bibinfo {author} {\bibfnamefont {J.}~\bibnamefont {Ebad-Allah}}, \bibinfo
  {author} {\bibfnamefont {F.}~\bibnamefont {Freund}}, \bibinfo {author}
  {\bibfnamefont {A.}~\bibnamefont {Jesche}}, \bibinfo {author} {\bibfnamefont
  {A.~A.}\ \bibnamefont {Tsirlin}}, \bibinfo {author} {\bibfnamefont
  {M.}~\bibnamefont {Hanfland}}, \bibinfo {author} {\bibfnamefont
  {P.}~\bibnamefont {Gegenwart}}, \bibinfo {author} {\bibfnamefont {I.~I.}\
  \bibnamefont {Mazin}}, \bibinfo {author} {\bibfnamefont {D.~I.}\ \bibnamefont
  {Khomskii}}, \bibinfo {author} {\bibfnamefont {R.}~\bibnamefont
  {Valent\'\i}}, \ and\ \bibinfo {author} {\bibfnamefont {C.~A.}\ \bibnamefont
  {Kuntscher}},\ }\bibfield  {title} {\enquote {\bibinfo {title} {{Competition
  between spin-orbit coupling, magnetism, and dimerization in the honeycomb
  iridates: $\upalpha$-Li\textsubscript{2}IrO\textsubscript{3} under
  pressure}},}\ }\href {\doibase 10.1103/PhysRevB.97.020104} {\bibfield
  {journal} {\bibinfo  {journal} {{Phys. Rev. B}}\ }\textbf {\bibinfo {volume}
  {97}},\ \bibinfo {pages} {020104(R)} (\bibinfo {year} {2018})}\BibitemShut
  {NoStop}%
\bibitem [{\citenamefont {Bastien}\ \emph {et~al.}(2018)\citenamefont
  {Bastien}, \citenamefont {Garbarino}, \citenamefont {Yadav}, \citenamefont
  {Martinez-Casado}, \citenamefont {Beltr\'an~Rodr\'{\i}guez}, \citenamefont
  {Stahl}, \citenamefont {Kusch}, \citenamefont {Limandri}, \citenamefont
  {Ray}, \citenamefont {Lampen-Kelley}, \citenamefont {Mandrus}, \citenamefont
  {Nagler}, \citenamefont {Roslova}, \citenamefont {Isaeva}, \citenamefont
  {Doert}, \citenamefont {Hozoi}, \citenamefont {Wolter}, \citenamefont
  {B\"uchner}, \citenamefont {Geck},\ and\ \citenamefont {van~den
  Brink}}]{Bastien.2018}%
  \BibitemOpen
  \bibfield  {author} {\bibinfo {author} {\bibfnamefont {G.}~\bibnamefont
  {Bastien}}, \bibinfo {author} {\bibfnamefont {G.}~\bibnamefont {Garbarino}},
  \bibinfo {author} {\bibfnamefont {R.}~\bibnamefont {Yadav}}, \bibinfo
  {author} {\bibfnamefont {F.~J.}\ \bibnamefont {Martinez-Casado}}, \bibinfo
  {author} {\bibfnamefont {R.}~\bibnamefont {Beltr\'an~Rodr\'{\i}guez}},
  \bibinfo {author} {\bibfnamefont {Q.}~\bibnamefont {Stahl}}, \bibinfo
  {author} {\bibfnamefont {M.}~\bibnamefont {Kusch}}, \bibinfo {author}
  {\bibfnamefont {S.~P.}\ \bibnamefont {Limandri}}, \bibinfo {author}
  {\bibfnamefont {R.}~\bibnamefont {Ray}}, \bibinfo {author} {\bibfnamefont
  {P.}~\bibnamefont {Lampen-Kelley}}, \bibinfo {author} {\bibfnamefont {D.~G.}\
  \bibnamefont {Mandrus}}, \bibinfo {author} {\bibfnamefont {S.~E.}\
  \bibnamefont {Nagler}}, \bibinfo {author} {\bibfnamefont {M.}~\bibnamefont
  {Roslova}}, \bibinfo {author} {\bibfnamefont {A.}~\bibnamefont {Isaeva}},
  \bibinfo {author} {\bibfnamefont {T.}~\bibnamefont {Doert}}, \bibinfo
  {author} {\bibfnamefont {L.}~\bibnamefont {Hozoi}}, \bibinfo {author}
  {\bibfnamefont {A.~U.~B.}\ \bibnamefont {Wolter}}, \bibinfo {author}
  {\bibfnamefont {B.}~\bibnamefont {B\"uchner}}, \bibinfo {author}
  {\bibfnamefont {J.}~\bibnamefont {Geck}}, \ and\ \bibinfo {author}
  {\bibfnamefont {J.}~\bibnamefont {van~den Brink}},\ }\bibfield  {title}
  {\enquote {\bibinfo {title} {{Pressure-induced dimerization and valence bond
  crystal formation in the Kitaev-Heisenberg magnet
  $\ensuremath{\alpha}\text{\ensuremath{-}}{\mathrm{RuCl}}_{3}$}},}\ }\href
  {\doibase 10.1103/PhysRevB.97.241108} {\bibfield  {journal} {\bibinfo
  {journal} {Phys. Rev. B}\ }\textbf {\bibinfo {volume} {97}},\ \bibinfo
  {pages} {241108} (\bibinfo {year} {2018})}\BibitemShut {NoStop}%
\bibitem [{\citenamefont {Hu}\ \emph {et~al.}(2018)\citenamefont {Hu},
  \citenamefont {Zhou}, \citenamefont {Wei}, \citenamefont {Li},\ and\
  \citenamefont {Feng}}]{Hu.2018}%
  \BibitemOpen
  \bibfield  {author} {\bibinfo {author} {\bibfnamefont {K.}~\bibnamefont
  {Hu}}, \bibinfo {author} {\bibfnamefont {Z.}~\bibnamefont {Zhou}}, \bibinfo
  {author} {\bibfnamefont {Y.-W.}\ \bibnamefont {Wei}}, \bibinfo {author}
  {\bibfnamefont {C.-K.}\ \bibnamefont {Li}}, \ and\ \bibinfo {author}
  {\bibfnamefont {J.}~\bibnamefont {Feng}},\ }\bibfield  {title} {\enquote
  {\bibinfo {title} {{Bond ordering and phase transitions in
  ${\mathrm{Na}}_{2}{\mathrm{IrO}}_{3}$ under high pressure}},}\ }\href
  {\doibase 10.1103/PhysRevB.98.100103} {\bibfield  {journal} {\bibinfo
  {journal} {Phys. Rev. B}\ }\textbf {\bibinfo {volume} {98}},\ \bibinfo
  {pages} {100103} (\bibinfo {year} {2018})}\BibitemShut {NoStop}%
\bibitem [{\citenamefont {Majumder}\ \emph {et~al.}(2018)\citenamefont
  {Majumder}, \citenamefont {Manna}, \citenamefont {Simutis}, \citenamefont
  {Orain}, \citenamefont {Dey}, \citenamefont {Freund}, \citenamefont {Jesche},
  \citenamefont {Khasanov}, \citenamefont {Biswas}, \citenamefont {Bykova},
  \citenamefont {Dubrovinskaia}, \citenamefont {Dubrovinsky}, \citenamefont
  {Yadav}, \citenamefont {Hozoi}, \citenamefont {Nishimoto}, \citenamefont
  {Tsirlin},\ and\ \citenamefont {Gegenwart}}]{Majumder.2018}%
  \BibitemOpen
  \bibfield  {author} {\bibinfo {author} {\bibfnamefont {M.}~\bibnamefont
  {Majumder}}, \bibinfo {author} {\bibfnamefont {R.~S.}\ \bibnamefont {Manna}},
  \bibinfo {author} {\bibfnamefont {G.}~\bibnamefont {Simutis}}, \bibinfo
  {author} {\bibfnamefont {J.~C.}\ \bibnamefont {Orain}}, \bibinfo {author}
  {\bibfnamefont {T.}~\bibnamefont {Dey}}, \bibinfo {author} {\bibfnamefont
  {F.}~\bibnamefont {Freund}}, \bibinfo {author} {\bibfnamefont
  {A.}~\bibnamefont {Jesche}}, \bibinfo {author} {\bibfnamefont
  {R.}~\bibnamefont {Khasanov}}, \bibinfo {author} {\bibfnamefont {P.~K.}\
  \bibnamefont {Biswas}}, \bibinfo {author} {\bibfnamefont {E.}~\bibnamefont
  {Bykova}}, \bibinfo {author} {\bibfnamefont {N.}~\bibnamefont
  {Dubrovinskaia}}, \bibinfo {author} {\bibfnamefont {L.~S.}\ \bibnamefont
  {Dubrovinsky}}, \bibinfo {author} {\bibfnamefont {R.}~\bibnamefont {Yadav}},
  \bibinfo {author} {\bibfnamefont {L.}~\bibnamefont {Hozoi}}, \bibinfo
  {author} {\bibfnamefont {S.}~\bibnamefont {Nishimoto}}, \bibinfo {author}
  {\bibfnamefont {A.~A.}\ \bibnamefont {Tsirlin}}, \ and\ \bibinfo {author}
  {\bibfnamefont {P.}~\bibnamefont {Gegenwart}},\ }\bibfield  {title} {\enquote
  {\bibinfo {title} {{Breakdown of Magnetic Order in the Pressurized Kitaev
  Iridate
  $\ensuremath{\beta}\text{\ensuremath{-}}{\mathrm{Li}}_{2}{\mathrm{IrO}}_{3}$}},}\
  }\href {\doibase 10.1103/PhysRevLett.120.237202} {\bibfield  {journal}
  {\bibinfo  {journal} {Phys. Rev. Lett.}\ }\textbf {\bibinfo {volume} {120}},\
  \bibinfo {pages} {237202} (\bibinfo {year} {2018})}\BibitemShut {NoStop}%
\bibitem [{1()}]{1}%
  \BibitemOpen
  \href@noop {} {}\bibinfo {note} {For Na$_2$IrO$_3$, additional
  pressure-induced transitions were suggested recently from powder x-ray
  diffraction measurements~\cite {Xi.2018}, which were, however, not observed
  by higher-quality single-crystal x-ray diffraction data~\cite
  {Hermann.2017}.}\BibitemShut {Stop}%
\bibitem [{\citenamefont {Yadav}\ \emph {et~al.}(2018)\citenamefont {Yadav},
  \citenamefont {Rachel}, \citenamefont {Hozoi}, \citenamefont {van~den
  Brink},\ and\ \citenamefont {Jackeli}}]{Yadav.2018}%
  \BibitemOpen
  \bibfield  {author} {\bibinfo {author} {\bibfnamefont {R.}~\bibnamefont
  {Yadav}}, \bibinfo {author} {\bibfnamefont {S.}~\bibnamefont {Rachel}},
  \bibinfo {author} {\bibfnamefont {L.}~\bibnamefont {Hozoi}}, \bibinfo
  {author} {\bibfnamefont {J.}~\bibnamefont {van~den Brink}}, \ and\ \bibinfo
  {author} {\bibfnamefont {G.}~\bibnamefont {Jackeli}},\ }\bibfield  {title}
  {\enquote {\bibinfo {title} {Strain- and pressure-tuned magnetic interactions
  in honeycomb kitaev materials},}\ }\href {\doibase
  10.1103/PhysRevB.98.121107} {\bibfield  {journal} {\bibinfo  {journal} {Phys.
  Rev. B}\ }\textbf {\bibinfo {volume} {98}},\ \bibinfo {pages} {121107}
  (\bibinfo {year} {2018})}\BibitemShut {NoStop}%
\bibitem [{\citenamefont {Gretarsson}\ \emph {et~al.}(2013)\citenamefont
  {Gretarsson}, \citenamefont {Clancy}, \citenamefont {Liu}, \citenamefont
  {Hill}, \citenamefont {Bozin}, \citenamefont {Singh}, \citenamefont {Manni},
  \citenamefont {Gegenwart}, \citenamefont {Kim}, \citenamefont {Said},
  \citenamefont {Casa}, \citenamefont {Gog}, \citenamefont {Upton},
  \citenamefont {Kim}, \citenamefont {Yu}, \citenamefont {Katukuri},
  \citenamefont {Hozoi}, \citenamefont {{van den Brink}},\ and\ \citenamefont
  {Kim}}]{Gretarsson.2013}%
  \BibitemOpen
  \bibfield  {author} {\bibinfo {author} {\bibfnamefont {H.}~\bibnamefont
  {Gretarsson}}, \bibinfo {author} {\bibfnamefont {J.~P.}\ \bibnamefont
  {Clancy}}, \bibinfo {author} {\bibfnamefont {X.}~\bibnamefont {Liu}},
  \bibinfo {author} {\bibfnamefont {J.~P.}\ \bibnamefont {Hill}}, \bibinfo
  {author} {\bibfnamefont {E.}~\bibnamefont {Bozin}}, \bibinfo {author}
  {\bibfnamefont {Y.}~\bibnamefont {Singh}}, \bibinfo {author} {\bibfnamefont
  {S.}~\bibnamefont {Manni}}, \bibinfo {author} {\bibfnamefont
  {P.}~\bibnamefont {Gegenwart}}, \bibinfo {author} {\bibfnamefont
  {J.}~\bibnamefont {Kim}}, \bibinfo {author} {\bibfnamefont {A.~H.}\
  \bibnamefont {Said}}, \bibinfo {author} {\bibfnamefont {D.}~\bibnamefont
  {Casa}}, \bibinfo {author} {\bibfnamefont {T.}~\bibnamefont {Gog}}, \bibinfo
  {author} {\bibfnamefont {M.~H.}\ \bibnamefont {Upton}}, \bibinfo {author}
  {\bibfnamefont {H.-S.}\ \bibnamefont {Kim}}, \bibinfo {author} {\bibfnamefont
  {J.}~\bibnamefont {Yu}}, \bibinfo {author} {\bibfnamefont {V.~M.}\
  \bibnamefont {Katukuri}}, \bibinfo {author} {\bibfnamefont {L.}~\bibnamefont
  {Hozoi}}, \bibinfo {author} {\bibfnamefont {J.}~\bibnamefont {{van den
  Brink}}}, \ and\ \bibinfo {author} {\bibfnamefont {Y.-J.}\ \bibnamefont
  {Kim}},\ }\bibfield  {title} {\enquote {\bibinfo {title} {{Crystal-Field
  Splitting and Correlation Effect on the Electronic Structure of
  \textit{A}\textsubscript{2}IrO\textsubscript{3}}},}\ }\href {\doibase
  10.1103/PhysRevLett.110.076402} {\bibfield  {journal} {\bibinfo  {journal}
  {{Phys. Rev. Lett.}}\ }\textbf {\bibinfo {volume} {110}},\ \bibinfo {pages}
  {076402} (\bibinfo {year} {2013})}\BibitemShut {NoStop}%
\bibitem [{\citenamefont {Kim}\ \emph {et~al.}(2012)\citenamefont {Kim},
  \citenamefont {Kim}, \citenamefont {Jeong}, \citenamefont {Jin},\ and\
  \citenamefont {Yu}}]{Kim.2012}%
  \BibitemOpen
  \bibfield  {author} {\bibinfo {author} {\bibfnamefont {C.~H.}\ \bibnamefont
  {Kim}}, \bibinfo {author} {\bibfnamefont {H.~S.}\ \bibnamefont {Kim}},
  \bibinfo {author} {\bibfnamefont {H.}~\bibnamefont {Jeong}}, \bibinfo
  {author} {\bibfnamefont {H.}~\bibnamefont {Jin}}, \ and\ \bibinfo {author}
  {\bibfnamefont {J.}~\bibnamefont {Yu}},\ }\bibfield  {title} {\enquote
  {\bibinfo {title} {{Topological Quantum Phase Transition in 5d Transition
  Metal Oxide Na\textsubscript{2}IrO\textsubscript{3}}},}\ }\href {\doibase
  10.1103/PhysRevLett.108.106401} {\bibfield  {journal} {\bibinfo  {journal}
  {{Phys. Rev. Lett.}}\ }\textbf {\bibinfo {volume} {108}},\ \bibinfo {pages}
  {106401} (\bibinfo {year} {2012})}\BibitemShut {NoStop}%
\bibitem [{\citenamefont {Kim}\ \emph {et~al.}(2013)\citenamefont {Kim},
  \citenamefont {Kim}, \citenamefont {Jeong}, \citenamefont {Jin},\ and\
  \citenamefont {Yu}}]{Kim.2013}%
  \BibitemOpen
  \bibfield  {author} {\bibinfo {author} {\bibfnamefont {H.-S.}\ \bibnamefont
  {Kim}}, \bibinfo {author} {\bibfnamefont {C.~H.}\ \bibnamefont {Kim}},
  \bibinfo {author} {\bibfnamefont {H.}~\bibnamefont {Jeong}}, \bibinfo
  {author} {\bibfnamefont {H.}~\bibnamefont {Jin}}, \ and\ \bibinfo {author}
  {\bibfnamefont {J.}~\bibnamefont {Yu}},\ }\bibfield  {title} {\enquote
  {\bibinfo {title} {{Strain-induced topological insulator phase and effective
  magnetic interactions in Li\textsubscript{2}IrO\textsubscript{3}}},}\ }\href
  {\doibase 10.1103/PhysRevB.87.165117} {\bibfield  {journal} {\bibinfo
  {journal} {{Phys. Rev. B}}\ }\textbf {\bibinfo {volume} {87}},\ \bibinfo
  {pages} {165117} (\bibinfo {year} {2013})}\BibitemShut {NoStop}%
\bibitem [{\citenamefont {Mazin}\ \emph {et~al.}(2012)\citenamefont {Mazin},
  \citenamefont {Jeschke}, \citenamefont {Foyevtsova}, \citenamefont
  {Valent\'\i},\ and\ \citenamefont {Khomskii}}]{Mazin.2012}%
  \BibitemOpen
  \bibfield  {author} {\bibinfo {author} {\bibfnamefont {I.~I.}\ \bibnamefont
  {Mazin}}, \bibinfo {author} {\bibfnamefont {H.~O.}\ \bibnamefont {Jeschke}},
  \bibinfo {author} {\bibfnamefont {K.}~\bibnamefont {Foyevtsova}}, \bibinfo
  {author} {\bibfnamefont {R.}~\bibnamefont {Valent\'\i}}, \ and\ \bibinfo
  {author} {\bibfnamefont {D.~I.}\ \bibnamefont {Khomskii}},\ }\bibfield
  {title} {\enquote {\bibinfo {title} {{Na\textsubscript{2}IrO\textsubscript{3}
  as a Molecular Orbital Crystal}},}\ }\href {\doibase
  10.1103/PhysRevLett.109.197201} {\bibfield  {journal} {\bibinfo  {journal}
  {{Phys. Rev. Lett.}}\ }\textbf {\bibinfo {volume} {109}},\ \bibinfo {pages}
  {197201} (\bibinfo {year} {2012})}\BibitemShut {NoStop}%
\bibitem [{\citenamefont {Mazin}\ \emph {et~al.}(2013)\citenamefont {Mazin},
  \citenamefont {Manni}, \citenamefont {Foyevtsova}, \citenamefont {Jeschke},
  \citenamefont {Gegenwart},\ and\ \citenamefont {Valent\'\i}}]{Mazin.2013}%
  \BibitemOpen
  \bibfield  {author} {\bibinfo {author} {\bibfnamefont {I.~I.}\ \bibnamefont
  {Mazin}}, \bibinfo {author} {\bibfnamefont {S.}~\bibnamefont {Manni}},
  \bibinfo {author} {\bibfnamefont {K.}~\bibnamefont {Foyevtsova}}, \bibinfo
  {author} {\bibfnamefont {H.~O.}\ \bibnamefont {Jeschke}}, \bibinfo {author}
  {\bibfnamefont {P.}~\bibnamefont {Gegenwart}}, \ and\ \bibinfo {author}
  {\bibfnamefont {R.}~\bibnamefont {Valent\'\i}},\ }\bibfield  {title}
  {\enquote {\bibinfo {title} {{Origin of the insulating state in honeycomb
  iridates and rhodates}},}\ }\href {\doibase 10.1103/PhysRevB.88.035115}
  {\bibfield  {journal} {\bibinfo  {journal} {{Phys. Rev. B}}\ }\textbf
  {\bibinfo {volume} {88}},\ \bibinfo {pages} {035115} (\bibinfo {year}
  {2013})}\BibitemShut {NoStop}%
\bibitem [{\citenamefont {Singh}\ and\ \citenamefont
  {Gegenwart}(2010)}]{Singh.2010}%
  \BibitemOpen
  \bibfield  {author} {\bibinfo {author} {\bibfnamefont {Y.}~\bibnamefont
  {Singh}}\ and\ \bibinfo {author} {\bibfnamefont {P.}~\bibnamefont
  {Gegenwart}},\ }\bibfield  {title} {\enquote {\bibinfo {title}
  {{Antiferromagnetic Mott insulating state in single crystals of the honeycomb
  lattice material Na\textsubscript{2}IrO\textsubscript{3}}},}\ }\href
  {\doibase 10.1103/PhysRevB.82.064412} {\bibfield  {journal} {\bibinfo
  {journal} {{Phys. Rev. B}}\ }\textbf {\bibinfo {volume} {82}},\ \bibinfo
  {pages} {064412} (\bibinfo {year} {2010})}\BibitemShut {NoStop}%
\bibitem [{\citenamefont {Xi}\ \emph {et~al.}(2018)\citenamefont {Xi},
  \citenamefont {Bo}, \citenamefont {Xu}, \citenamefont {Kong}, \citenamefont
  {Liu}, \citenamefont {Hong}, \citenamefont {Jin}, \citenamefont {Cao},
  \citenamefont {Wan},\ and\ \citenamefont {Carr}}]{Xi.2018}%
  \BibitemOpen
  \bibfield  {author} {\bibinfo {author} {\bibfnamefont {X.}~\bibnamefont
  {Xi}}, \bibinfo {author} {\bibfnamefont {X.}~\bibnamefont {Bo}}, \bibinfo
  {author} {\bibfnamefont {X.~S.}\ \bibnamefont {Xu}}, \bibinfo {author}
  {\bibfnamefont {P.~P.}\ \bibnamefont {Kong}}, \bibinfo {author}
  {\bibfnamefont {Z.}~\bibnamefont {Liu}}, \bibinfo {author} {\bibfnamefont
  {X.~G.}\ \bibnamefont {Hong}}, \bibinfo {author} {\bibfnamefont {C.~Q.}\
  \bibnamefont {Jin}}, \bibinfo {author} {\bibfnamefont {G.}~\bibnamefont
  {Cao}}, \bibinfo {author} {\bibfnamefont {X.}~\bibnamefont {Wan}}, \ and\
  \bibinfo {author} {\bibfnamefont {G.~L.}\ \bibnamefont {Carr}},\ }\bibfield
  {title} {\enquote {\bibinfo {title} {{Honeycomb lattice Na$_2$IrO$_3$ at high
  pressures: A robust spin-orbit Mott insulator}},}\ }\href {\doibase
  10.1103/PhysRevB.98.125117} {\bibfield  {journal} {\bibinfo  {journal}
  {{Phys. Rev. B}}\ }\textbf {\bibinfo {volume} {98}},\ \bibinfo {pages}
  {125117} (\bibinfo {year} {2018})}\BibitemShut {NoStop}%
\bibitem [{\citenamefont {Sohn}\ \emph {et~al.}(2013)\citenamefont {Sohn},
  \citenamefont {Kim}, \citenamefont {Qi}, \citenamefont {Jeong}, \citenamefont
  {Park}, \citenamefont {Yoo}, \citenamefont {Kim}, \citenamefont {Kim},
  \citenamefont {Kang}, \citenamefont {Cho}, \citenamefont {Cao}, \citenamefont
  {Yu}, \citenamefont {Moon},\ and\ \citenamefont {Noh}}]{Sohn.2013}%
  \BibitemOpen
  \bibfield  {author} {\bibinfo {author} {\bibfnamefont {C.~H.}\ \bibnamefont
  {Sohn}}, \bibinfo {author} {\bibfnamefont {H.-S.}\ \bibnamefont {Kim}},
  \bibinfo {author} {\bibfnamefont {T.~F.}\ \bibnamefont {Qi}}, \bibinfo
  {author} {\bibfnamefont {D.~W.}\ \bibnamefont {Jeong}}, \bibinfo {author}
  {\bibfnamefont {H.~J.}\ \bibnamefont {Park}}, \bibinfo {author}
  {\bibfnamefont {H.~K.}\ \bibnamefont {Yoo}}, \bibinfo {author} {\bibfnamefont
  {H.~H.}\ \bibnamefont {Kim}}, \bibinfo {author} {\bibfnamefont {J.-Y.}\
  \bibnamefont {Kim}}, \bibinfo {author} {\bibfnamefont {T.~D.}\ \bibnamefont
  {Kang}}, \bibinfo {author} {\bibfnamefont {D.-Y.}\ \bibnamefont {Cho}},
  \bibinfo {author} {\bibfnamefont {G.}~\bibnamefont {Cao}}, \bibinfo {author}
  {\bibfnamefont {J.}~\bibnamefont {Yu}}, \bibinfo {author} {\bibfnamefont
  {S.~J.}\ \bibnamefont {Moon}}, \ and\ \bibinfo {author} {\bibfnamefont
  {T.~W.}\ \bibnamefont {Noh}},\ }\bibfield  {title} {\enquote {\bibinfo
  {title} {{Mixing between $J_\text{eff}=1/2$ and $3/2$ orbitals in
  Na\textsubscript{2}IrO\textsubscript{3}: A spectroscopic and density
  functional calculation study}},}\ }\href {\doibase
  10.1103/PhysRevB.88.085125} {\bibfield  {journal} {\bibinfo  {journal}
  {{Phys. Rev. B}}\ }\textbf {\bibinfo {volume} {88}},\ \bibinfo {pages}
  {085125} (\bibinfo {year} {2013})}\BibitemShut {NoStop}%
\bibitem [{\citenamefont {Kim}\ \emph {et~al.}(2016)\citenamefont {Kim},
  \citenamefont {Shirakawa},\ and\ \citenamefont {Yunoki}}]{Kim.2016c}%
  \BibitemOpen
  \bibfield  {author} {\bibinfo {author} {\bibfnamefont {B.~H.}\ \bibnamefont
  {Kim}}, \bibinfo {author} {\bibfnamefont {T.}~\bibnamefont {Shirakawa}}, \
  and\ \bibinfo {author} {\bibfnamefont {S.}~\bibnamefont {Yunoki}},\
  }\bibfield  {title} {\enquote {\bibinfo {title} {{From a Quasimolecular Band
  Insulator to a Relativistic Mott Insulator in $t_{2g}^5$ Systems with a
  Honeycomb Lattice Structure}},}\ }\href {\doibase
  10.1103/PhysRevLett.117.187201} {\bibfield  {journal} {\bibinfo  {journal}
  {{Phys. Rev. Lett.}}\ }\textbf {\bibinfo {volume} {117}},\ \bibinfo {pages}
  {187201} (\bibinfo {year} {2016})}\BibitemShut {NoStop}%
\bibitem [{\citenamefont {Clancy}\ \emph {et~al.}(2018)\citenamefont {Clancy},
  \citenamefont {Gretarsson}, \citenamefont {Sears}, \citenamefont {Singh},
  \citenamefont {Desgreniers}, \citenamefont {Mehlawat}, \citenamefont {Layek},
  \citenamefont {Rozenberg}, \citenamefont {Ding}, \citenamefont {Upton},
  \citenamefont {Casa}, \citenamefont {Chen}, \citenamefont {Im}, \citenamefont
  {Lee}, \citenamefont {Yadav}, \citenamefont {Hozoi}, \citenamefont {Efremov},
  \citenamefont {{van den Brink}},\ and\ \citenamefont {Kim}}]{Clancy.2018}%
  \BibitemOpen
  \bibfield  {author} {\bibinfo {author} {\bibfnamefont {J.~P.}\ \bibnamefont
  {Clancy}}, \bibinfo {author} {\bibfnamefont {H.}~\bibnamefont {Gretarsson}},
  \bibinfo {author} {\bibfnamefont {J.~A.}\ \bibnamefont {Sears}}, \bibinfo
  {author} {\bibfnamefont {Y.}~\bibnamefont {Singh}}, \bibinfo {author}
  {\bibfnamefont {S.}~\bibnamefont {Desgreniers}}, \bibinfo {author}
  {\bibfnamefont {K.}~\bibnamefont {Mehlawat}}, \bibinfo {author}
  {\bibfnamefont {S.}~\bibnamefont {Layek}}, \bibinfo {author} {\bibfnamefont
  {G.~Kh.}\ \bibnamefont {Rozenberg}}, \bibinfo {author} {\bibfnamefont
  {Y.}~\bibnamefont {Ding}}, \bibinfo {author} {\bibfnamefont {M.~H.}\
  \bibnamefont {Upton}}, \bibinfo {author} {\bibfnamefont {D.}~\bibnamefont
  {Casa}}, \bibinfo {author} {\bibfnamefont {N.}~\bibnamefont {Chen}}, \bibinfo
  {author} {\bibfnamefont {J.}~\bibnamefont {Im}}, \bibinfo {author}
  {\bibfnamefont {Y.}~\bibnamefont {Lee}}, \bibinfo {author} {\bibfnamefont
  {R.}~\bibnamefont {Yadav}}, \bibinfo {author} {\bibfnamefont
  {L.}~\bibnamefont {Hozoi}}, \bibinfo {author} {\bibfnamefont
  {D.}~\bibnamefont {Efremov}}, \bibinfo {author} {\bibfnamefont
  {J.}~\bibnamefont {{van den Brink}}}, \ and\ \bibinfo {author} {\bibfnamefont
  {Y.-J.}\ \bibnamefont {Kim}},\ }\bibfield  {title} {\enquote {\bibinfo
  {title} {{Pressure-driven collapse of the relativistic electronic ground
  state in a honeycomb iridate}},}\ }\href {\doibase 10.1038/s41535-018-0109-0}
  {\bibfield  {journal} {\bibinfo  {journal} {{npj Quantum Mater.}}\ }\textbf
  {\bibinfo {volume} {3}},\ \bibinfo {pages} {57} (\bibinfo {year}
  {2018})}\BibitemShut {NoStop}%
\bibitem [{\citenamefont {Foyevtsova}\ \emph {et~al.}(2013)\citenamefont
  {Foyevtsova}, \citenamefont {Jeschke}, \citenamefont {Mazin}, \citenamefont
  {Khomskii},\ and\ \citenamefont {Valent\'\i}}]{Foyevtsova.2013}%
  \BibitemOpen
  \bibfield  {author} {\bibinfo {author} {\bibfnamefont {K.}~\bibnamefont
  {Foyevtsova}}, \bibinfo {author} {\bibfnamefont {H.~O.}\ \bibnamefont
  {Jeschke}}, \bibinfo {author} {\bibfnamefont {I.~I.}\ \bibnamefont {Mazin}},
  \bibinfo {author} {\bibfnamefont {D.~I.}\ \bibnamefont {Khomskii}}, \ and\
  \bibinfo {author} {\bibfnamefont {R.}~\bibnamefont {Valent\'\i}},\ }\bibfield
   {title} {\enquote {\bibinfo {title} {{Ab initio analysis of the
  tight-binding parameters and magnetic interactions in
  Na\textsubscript{2}IrO\textsubscript{3}}},}\ }\href {\doibase
  10.1103/PhysRevB.88.035107} {\bibfield  {journal} {\bibinfo  {journal}
  {{Phys. Rev. B}}\ }\textbf {\bibinfo {volume} {88}},\ \bibinfo {pages}
  {035107} (\bibinfo {year} {2013})}\BibitemShut {NoStop}%
\bibitem [{\citenamefont {Bhattacharjee}\ \emph {et~al.}(2012)\citenamefont
  {Bhattacharjee}, \citenamefont {Lee},\ and\ \citenamefont
  {Kim}}]{Bhattacharjee.2012}%
  \BibitemOpen
  \bibfield  {author} {\bibinfo {author} {\bibfnamefont {S.}~\bibnamefont
  {Bhattacharjee}}, \bibinfo {author} {\bibfnamefont {S.-S.}\ \bibnamefont
  {Lee}}, \ and\ \bibinfo {author} {\bibfnamefont {Y.~B.}\ \bibnamefont
  {Kim}},\ }\bibfield  {title} {\enquote {\bibinfo {title} {Spin–-orbital
  locking, emergent pseudo-spin and magnetic order in honeycomb lattice
  iridates},}\ }\href {\doibase 10.1088/1367-2630/14/7/073015} {\bibfield
  {journal} {\bibinfo  {journal} {{New J. Phys.}}\ }\textbf {\bibinfo {volume}
  {14}},\ \bibinfo {pages} {073015} (\bibinfo {year} {2012})}\BibitemShut
  {NoStop}%
\bibitem [{\citenamefont {Sun}\ \emph {et~al.}(2018)\citenamefont {Sun},
  \citenamefont {Zheng}, \citenamefont {Liu}, \citenamefont {Sandoval},
  \citenamefont {Xu}, \citenamefont {Xu}, \citenamefont {Jin}, \citenamefont
  {Sun}, \citenamefont {Yang}, \citenamefont {Mao}, \citenamefont {Mitchell},
  \citenamefont {Kolmogorov},\ and\ \citenamefont {Haskel}}]{Sun.2018}%
  \BibitemOpen
  \bibfield  {author} {\bibinfo {author} {\bibfnamefont {F.}~\bibnamefont
  {Sun}}, \bibinfo {author} {\bibfnamefont {H.}~\bibnamefont {Zheng}}, \bibinfo
  {author} {\bibfnamefont {Y.}~\bibnamefont {Liu}}, \bibinfo {author}
  {\bibfnamefont {E.~D.}\ \bibnamefont {Sandoval}}, \bibinfo {author}
  {\bibfnamefont {C.}~\bibnamefont {Xu}}, \bibinfo {author} {\bibfnamefont
  {J.}~\bibnamefont {Xu}}, \bibinfo {author} {\bibfnamefont {C.~Q.}\
  \bibnamefont {Jin}}, \bibinfo {author} {\bibfnamefont {C.~J.}\ \bibnamefont
  {Sun}}, \bibinfo {author} {\bibfnamefont {W.~G.}\ \bibnamefont {Yang}},
  \bibinfo {author} {\bibfnamefont {H.~K.}\ \bibnamefont {Mao}}, \bibinfo
  {author} {\bibfnamefont {J.~F.}\ \bibnamefont {Mitchell}}, \bibinfo {author}
  {\bibfnamefont {A.~N.}\ \bibnamefont {Kolmogorov}}, \ and\ \bibinfo {author}
  {\bibfnamefont {D.}~\bibnamefont {Haskel}},\ }\bibfield  {title} {\enquote
  {\bibinfo {title} {{Electronic and structural response to pressure in the
  hyperkagome-lattice ${\mathrm{Na}}_{3}{\mathrm{Ir}}_{3}{\mathrm{O}}_{8}$}},}\
  }\href {\doibase 10.1103/PhysRevB.98.085131} {\bibfield  {journal} {\bibinfo
  {journal} {Phys. Rev. B}\ }\textbf {\bibinfo {volume} {98}},\ \bibinfo
  {pages} {085131} (\bibinfo {year} {2018})}\BibitemShut {NoStop}%
\bibitem [{\citenamefont {Freund}\ \emph {et~al.}(2016)\citenamefont {Freund},
  \citenamefont {Williams}, \citenamefont {Johnson}, \citenamefont {Coldea},
  \citenamefont {Gegenwart},\ and\ \citenamefont {Jesche}}]{Freund.2016}%
  \BibitemOpen
  \bibfield  {author} {\bibinfo {author} {\bibfnamefont {F.}~\bibnamefont
  {Freund}}, \bibinfo {author} {\bibfnamefont {S.~C.}\ \bibnamefont
  {Williams}}, \bibinfo {author} {\bibfnamefont {R.~D.}\ \bibnamefont
  {Johnson}}, \bibinfo {author} {\bibfnamefont {R.}~\bibnamefont {Coldea}},
  \bibinfo {author} {\bibfnamefont {P.}~\bibnamefont {Gegenwart}}, \ and\
  \bibinfo {author} {\bibfnamefont {A.}~\bibnamefont {Jesche}},\ }\bibfield
  {title} {\enquote {\bibinfo {title} {{Single crystal growth from separated
  educts and its application to lithium transition-metal oxides}},}\ }\href
  {\doibase 10.1038/srep35362} {\bibfield  {journal} {\bibinfo  {journal}
  {{Sci. Rep.}}\ }\textbf {\bibinfo {volume} {6}},\ \bibinfo {pages} {35362}
  (\bibinfo {year} {2016})}\BibitemShut {NoStop}%
\bibitem [{\citenamefont {Huber}\ \emph {et~al.}(1977)\citenamefont {Huber},
  \citenamefont {Syassen},\ and\ \citenamefont {Holzapfel}}]{Huber.1977}%
  \BibitemOpen
  \bibfield  {author} {\bibinfo {author} {\bibfnamefont {G.}~\bibnamefont
  {Huber}}, \bibinfo {author} {\bibfnamefont {K.}~\bibnamefont {Syassen}}, \
  and\ \bibinfo {author} {\bibfnamefont {W.~B.}\ \bibnamefont {Holzapfel}},\
  }\bibfield  {title} {\enquote {\bibinfo {title} {{Pressure dependence of
  4\textit{f} levels in europium pentaphosphate up to 400~kbar}},}\ }\href
  {\doibase 10.1103/PhysRevB.15.5123} {\bibfield  {journal} {\bibinfo
  {journal} {{Phys. Rev. B}}\ }\textbf {\bibinfo {volume} {15}},\ \bibinfo
  {pages} {5123} (\bibinfo {year} {1977})}\BibitemShut {NoStop}%
\bibitem [{\citenamefont {Mao}\ \emph {et~al.}(1986)\citenamefont {Mao},
  \citenamefont {Xu},\ and\ \citenamefont {Bell}}]{Mao.1986}%
  \BibitemOpen
  \bibfield  {author} {\bibinfo {author} {\bibfnamefont {H.~K.}\ \bibnamefont
  {Mao}}, \bibinfo {author} {\bibfnamefont {J.}~\bibnamefont {Xu}}, \ and\
  \bibinfo {author} {\bibfnamefont {P.~M.}\ \bibnamefont {Bell}},\ }\bibfield
  {title} {\enquote {\bibinfo {title} {{Calibration of the ruby pressure gauge
  to 800 kbar under quasi‐hydrostatic conditions}},}\ }\href {\doibase
  10.1029/JB091iB05p04673} {\bibfield  {journal} {\bibinfo  {journal} {{J.
  Geoph. Re.: Solid Earth}}\ }\textbf {\bibinfo {volume} {91}},\ \bibinfo
  {pages} {4673} (\bibinfo {year} {1986})}\BibitemShut {NoStop}%
\bibitem [{\citenamefont {Eremets}\ and\ \citenamefont
  {Timofeev}(1992)}]{Eremets.1992}%
  \BibitemOpen
  \bibfield  {author} {\bibinfo {author} {\bibfnamefont {M.~I.}\ \bibnamefont
  {Eremets}}\ and\ \bibinfo {author} {\bibfnamefont {Yu.~A.}\ \bibnamefont
  {Timofeev}},\ }\bibfield  {title} {\enquote {\bibinfo {title} {{Miniature
  diamond anvil cell: Incorporating a new design for anvil alignment}},}\
  }\href {\doibase 10.1063/1.1143799} {\bibfield  {journal} {\bibinfo
  {journal} {{Rev. Sci. Instr.}}\ }\textbf {\bibinfo {volume} {63}},\ \bibinfo
  {pages} {3123} (\bibinfo {year} {1992})}\BibitemShut {NoStop}%
\bibitem [{\citenamefont {Ruoff}\ and\ \citenamefont
  {Ghandehari}(1994)}]{Ruoff.1994}%
  \BibitemOpen
  \bibfield  {author} {\bibinfo {author} {\bibfnamefont {A.~L.}\ \bibnamefont
  {Ruoff}}\ and\ \bibinfo {author} {\bibfnamefont {K.}~\bibnamefont
  {Ghandehari}},\ }\bibfield  {title} {\enquote {\bibinfo {title} {{Refractive
  index of diamond anvils at high pressure}},}\ }in\ \href {\doibase
  10.1063/1.46372} {\emph {\bibinfo {booktitle} {{AIP Conference
  Proceedings}}}}\ (\bibinfo  {publisher} {AIP},\ \bibinfo {year} {1994})\ pp.\
  \bibinfo {pages} {1523--1525}\BibitemShut {NoStop}%
\bibitem [{\citenamefont {Kuzmenko}(2005)}]{Kuzmenko.2005}%
  \BibitemOpen
  \bibfield  {author} {\bibinfo {author} {\bibfnamefont {A.~B.}\ \bibnamefont
  {Kuzmenko}},\ }\bibfield  {title} {\enquote {\bibinfo {title}
  {{Kramers--Kronig constrained variational analysis of optical spectra}},}\
  }\href {\doibase 10.1063/1.1979470} {\bibfield  {journal} {\bibinfo
  {journal} {{Rev. Sci. Instr.}}\ }\textbf {\bibinfo {volume} {76}},\ \bibinfo
  {pages} {083108} (\bibinfo {year} {2005})}\BibitemShut {NoStop}%
\bibitem [{\citenamefont {Plaskett}\ and\ \citenamefont
  {Schatz}(1963)}]{Plaskett.1963}%
  \BibitemOpen
  \bibfield  {author} {\bibinfo {author} {\bibfnamefont {J.~S.}\ \bibnamefont
  {Plaskett}}\ and\ \bibinfo {author} {\bibfnamefont {P.~N.}\ \bibnamefont
  {Schatz}},\ }\bibfield  {title} {\enquote {\bibinfo {title} {{On the Robinson
  and Price (Kramers--Kronig) Method of Interpreting Reflection Data Taken
  through a Transparent Window}},}\ }\href {\doibase 10.1063/1.1733714}
  {\bibfield  {journal} {\bibinfo  {journal} {{J. Chem. Phys.}}\ }\textbf
  {\bibinfo {volume} {38}},\ \bibinfo {pages} {612} (\bibinfo {year}
  {1963})}\BibitemShut {NoStop}%
\bibitem [{\citenamefont {Kroumova}\ \emph {et~al.}(2003)\citenamefont
  {Kroumova}, \citenamefont {Aroyo}, \citenamefont {Perez-Mato}, \citenamefont
  {Kirov}, \citenamefont {Capillas}, \citenamefont {Ivantchev},\ and\
  \citenamefont {Wondratschek}}]{Kroumova.2003}%
  \BibitemOpen
  \bibfield  {author} {\bibinfo {author} {\bibfnamefont {E.}~\bibnamefont
  {Kroumova}}, \bibinfo {author} {\bibfnamefont {M.~I.}\ \bibnamefont {Aroyo}},
  \bibinfo {author} {\bibfnamefont {J.~M.}\ \bibnamefont {Perez-Mato}},
  \bibinfo {author} {\bibfnamefont {A.}~\bibnamefont {Kirov}}, \bibinfo
  {author} {\bibfnamefont {C.}~\bibnamefont {Capillas}}, \bibinfo {author}
  {\bibfnamefont {S.}~\bibnamefont {Ivantchev}}, \ and\ \bibinfo {author}
  {\bibfnamefont {H.}~\bibnamefont {Wondratschek}},\ }\bibfield  {title}
  {\enquote {\bibinfo {title} {{Bilbao Crystallographic Server: Useful
  Databases and Tools for Phase-Transition Studies}},}\ }\href {\doibase
  10.1080/0141159031000076110} {\bibfield  {journal} {\bibinfo  {journal}
  {{Phase Transitions}}\ }\textbf {\bibinfo {volume} {76}},\ \bibinfo {pages}
  {155} (\bibinfo {year} {2003})}\BibitemShut {NoStop}%
\bibitem [{\citenamefont {Kim}\ \emph {et~al.}(2014)\citenamefont {Kim},
  \citenamefont {Khaliullin},\ and\ \citenamefont {Min}}]{Kim.2014}%
  \BibitemOpen
  \bibfield  {author} {\bibinfo {author} {\bibfnamefont {B.~H.}\ \bibnamefont
  {Kim}}, \bibinfo {author} {\bibfnamefont {G.}~\bibnamefont {Khaliullin}}, \
  and\ \bibinfo {author} {\bibfnamefont {B.~I.}\ \bibnamefont {Min}},\
  }\bibfield  {title} {\enquote {\bibinfo {title} {{Electronic excitations in
  the edge-shared relativistic Mott insulator:
  Na\textsubscript{2}IrO\textsubscript{3}}},}\ }\href {\doibase
  10.1103/PhysRevB.89.081109} {\bibfield  {journal} {\bibinfo  {journal}
  {{Phys. Rev. B}}\ }\textbf {\bibinfo {volume} {89}},\ \bibinfo {pages}
  {081109(R)} (\bibinfo {year} {2014})}\BibitemShut {NoStop}%
\bibitem [{\citenamefont {Li}\ \emph {et~al.}(2015)\citenamefont {Li},
  \citenamefont {Foyevtsova}, \citenamefont {Jeschke},\ and\ \citenamefont
  {Valent\'\i}}]{Li.2015}%
  \BibitemOpen
  \bibfield  {author} {\bibinfo {author} {\bibfnamefont {Y.}~\bibnamefont
  {Li}}, \bibinfo {author} {\bibfnamefont {K.}~\bibnamefont {Foyevtsova}},
  \bibinfo {author} {\bibfnamefont {H.~O.}\ \bibnamefont {Jeschke}}, \ and\
  \bibinfo {author} {\bibfnamefont {R.}~\bibnamefont {Valent\'\i}},\ }\bibfield
   {title} {\enquote {\bibinfo {title} {{Analysis of the optical conductivity
  for \textit{A}\textsubscript{2}IrO\textsubscript{3} (\textit{A}\,=\,Na, Li)
  from first principles}},}\ }\href {\doibase 10.1103/PhysRevB.91.161101}
  {\bibfield  {journal} {\bibinfo  {journal} {{Phys. Rev. B}}\ }\textbf
  {\bibinfo {volume} {91}},\ \bibinfo {pages} {161101} (\bibinfo {year}
  {2015})}\BibitemShut {NoStop}%
\bibitem [{\citenamefont {Li}\ \emph {et~al.}(2017)\citenamefont {Li},
  \citenamefont {Winter}, \citenamefont {Jeschke},\ and\ \citenamefont
  {Valent\'\i}}]{Li.2017}%
  \BibitemOpen
  \bibfield  {author} {\bibinfo {author} {\bibfnamefont {Ying}\ \bibnamefont
  {Li}}, \bibinfo {author} {\bibfnamefont {Stephen~M.}\ \bibnamefont {Winter}},
  \bibinfo {author} {\bibfnamefont {Harald~O.}\ \bibnamefont {Jeschke}}, \ and\
  \bibinfo {author} {\bibfnamefont {Roser}\ \bibnamefont {Valent\'\i}},\
  }\bibfield  {title} {\enquote {\bibinfo {title} {{Electronic excitations in
  $\upgamma$-Li\textsubscript{2}IrO\textsubscript{3}}},}\ }\href {\doibase
  10.1103/PhysRevB.95.045129} {\bibfield  {journal} {\bibinfo  {journal}
  {{Phys. Rev. B}}\ }\textbf {\bibinfo {volume} {95}},\ \bibinfo {pages}
  {045129} (\bibinfo {year} {2017})}\BibitemShut {NoStop}%
\bibitem [{Li.()}]{Li.2017b}%
  \BibitemOpen
  \href@noop {} {\enquote {\bibinfo {title} {{private communication AG
  Valent\'\i}},}\ }\BibitemShut {NoStop}%
\bibitem [{2()}]{2}%
  \BibitemOpen
  \href@noop {} {}\bibinfo {note} {We note that the contributions \textbf
  {A\textsubscript {2}} and \textbf {A\textsubscript {3}} would also fit
  energetically to interband transitions of QMOs, since $\upalpha
  $-Li$_2$IrO$_3$ is known to exhibit both Mott-insulating and QMO
  character~\cite {Kim.2016c,Hermann.2017} at ambient pressure.}\BibitemShut
  {Stop}%
\bibitem [{3()}]{3}%
  \BibitemOpen
  \href@noop {} {}\bibinfo {note} {A weakening of the optically forbidden
  on-site excitations may in general also be due to a reduction of the trigonal
  distortion of the IrO$_6$ octahedra~\cite
  {Mazin.2012,Bhattacharjee.2012,Kim.2012}. However, according to the energy
  difference $\Delta {A_{2,3}}$ [see Fig.\ \ref {fig.relA}(b)] the opposite
  trend is observed for $\alpha $-Li$_2$IrO$_3$, namely an increase in the
  trigonal distortion with increasing pressure up to $P\textsubscript
  {c}$.}\BibitemShut {Stop}%
\bibitem [{4()}]{4}%
  \BibitemOpen
  \href@noop {} {}\bibinfo {note} {According to the absorbance spectra [see
  Fig.\ \ref {fig.Absorbance}(c)] feature \textbf {A\textsubscript {2}}
  consists of two contributions \textbf {A$_\textbf {2}^\mathbf {\prime }$} and
  \textbf {A$_\textbf {2}^\mathbf {\prime \prime }$}.}\BibitemShut {Stop}%
\bibitem [{\citenamefont {Takayama}\ \emph {et~al.}()\citenamefont {Takayama},
  \citenamefont {Krajewska}, \citenamefont {Gibbs}, \citenamefont {Yaresko},
  \citenamefont {Ishii}, \citenamefont {Yamaoka}, \citenamefont {Ishii},
  \citenamefont {Hiraoka}, \citenamefont {Funnel}, \citenamefont {Bull},\ and\
  \citenamefont {Takagi}}]{Takayama.2018}%
  \BibitemOpen
  \bibfield  {author} {\bibinfo {author} {\bibfnamefont {T.}~\bibnamefont
  {Takayama}}, \bibinfo {author} {\bibfnamefont {A.}~\bibnamefont {Krajewska}},
  \bibinfo {author} {\bibfnamefont {A.~S.}\ \bibnamefont {Gibbs}}, \bibinfo
  {author} {\bibfnamefont {A.~N.}\ \bibnamefont {Yaresko}}, \bibinfo {author}
  {\bibfnamefont {H.}~\bibnamefont {Ishii}}, \bibinfo {author} {\bibfnamefont
  {H.}~\bibnamefont {Yamaoka}}, \bibinfo {author} {\bibfnamefont
  {K.}~\bibnamefont {Ishii}}, \bibinfo {author} {\bibfnamefont
  {N.}~\bibnamefont {Hiraoka}}, \bibinfo {author} {\bibfnamefont {N.~P.}\
  \bibnamefont {Funnel}}, \bibinfo {author} {\bibfnamefont {C.~L.}\
  \bibnamefont {Bull}}, \ and\ \bibinfo {author} {\bibfnamefont
  {H.}~\bibnamefont {Takagi}},\ }\href@noop {} {\enquote {\bibinfo {title}
  {{Pressure-induced collapse of spin-orbital Mott state in the hyperhoneycomb
  iridate $\upbeta$-Li\textsubscript{2}IrO\textsubscript{3}}},}\ }\Eprint
  {http://arxiv.org/abs/1808.05494v1} {arXiv:1808.05494v1} \BibitemShut
  {NoStop}%
\bibitem [{\citenamefont {Antonov}\ and\ \citenamefont
  {Uba}(2018)}]{Antonov.2018}%
  \BibitemOpen
  \bibfield  {author} {\bibinfo {author} {\bibfnamefont {V.~N.}\ \bibnamefont
  {Antonov}}\ and\ \bibinfo {author} {\bibfnamefont {L.}~\bibnamefont {Uba},
  \bibfnamefont {S.~und~Uba}},\ }\bibfield  {title} {\enquote {\bibinfo {title}
  {{Electronic structure and x-ray magnetic circular dichroism in the
  hyperhoneycomb iridate $\upbeta$-Li\textsubscript{2}IrO\textsubscript{3}}},}\
  }\href {\doibase 10.1103/PhysRevB.98.245113} {\bibfield  {journal} {\bibinfo
  {journal} {{Phys. Rev. B}}\ }\textbf {\bibinfo {volume} {98}},\ \bibinfo
  {pages} {245113} (\bibinfo {year} {2018})}\BibitemShut {NoStop}%
\end{thebibliography}
\end{document}